\documentclass[a4paper,11pt]{article}
\pdfoutput=1 

\usepackage{jheppub} 

\usepackage{graphicx}
\usepackage{dcolumn}
\usepackage{bm}
\usepackage{ulem}
\usepackage{float}
\usepackage{amsmath,amssymb,amsfonts}
\usepackage{todonotes}
\usepackage{multirow}
\usepackage{mathtools}
\usepackage{tikz}
\usepackage{soul}

\renewcommand{\bf}[1]{\mathbf{#1}}
\renewcommand{\emph}[1]{\textit{#1}}
\renewcommand{\d}{{\mathrm{d}}}
\newcommand{\U}{{\mathrm{U}}}

\renewcommand{\Im}{\mathrm{Im}}
\newcommand{\ar}{\vert \alpha_{\mathbf{k}}\rangle}
\newcommand{\al}{\langle\alpha_{\mathbf{k}} \vert}
\newcommand{\zr}{\vert 0 \rangle}
\newcommand{\zl}{\langle 0 \vert}
\newcommand{\p}{\mathbf{p}}
\renewcommand{\k}{\mathbf{k}}
\newcommand{\M}{\mathcal{M}}
\newcommand{\gap}{\Delta^2}
\newcommand{\be}{\begin{equation}}
\newcommand{\ee}{\end{equation}}

\usetikzlibrary{arrows}
\newcommand{\midarrow}{\tikz \draw[-triangle 90] (0,0) -- +(.1,0);}

\title{$S$-matrix positivity without Lorentz
invariance: a case study}
\author{Lam Hui,}
\author{Ioanna Kourkoulou,}
\author{Alberto Nicolis,}
\author{Alessandro Podo,}
\author{Shengjia Zhou}

\affiliation{Department of Physics, Center for Theoretical Physics, Columbia University, New York, 538 West 120th Street, NY 10027, USA}

\emailAdd{lh399@columbia.edu}
\emailAdd{ik2436@columbia.edu}
\emailAdd{a.nicolis@columbia.edu}
\emailAdd{ap3964@columbia.edu}
\emailAdd{sz2807@columbia.edu}

\abstract{We investigate the analytic structure of scattering amplitudes in theories in which Lorentz invariance is spontaneously broken. We do so by computing and studying the S-matrix for a simple example: a superfluid described by a complex scalar with quartic interactions. The computation is confined to tree-level, for there are no absolutely stable single-particle states, though the lifetime can be made long by lowering the chemical potential. For the $2 \to 2$ amplitude in center-of-mass configurations, not only is crossing symmetry violated, there appears a {\it tree level} branch cut for unphysical kinematics. Its appearance is a consequence of non-analyticity in the dispersion relation. The branch point defines a new scale in the problem, which scales inversely with the chemical potential. In this example, even derivatives of the forward amplitude are positive while odd derivatives are negative. This pattern can be understood in a general way in the limit of a small chemical potential, or weak Lorentz breaking.}

\begin{document}
\maketitle
\flushbottom

\section{Introduction}

Causality and analyticity can provide strong constraints on theoretical models of physical systems, as noticed long ago in the works of Kramers and Kr\"onig (see e.g.~\cite{Jackson:1998nia} for a modern account).
In Lorentz invariant quantum field theories (QFTs), these properties, together with unitarity and crossing symmetry, give rise to dispersive bounds on forward scattering amplitudes~\cite{Pham:1985cr,Adams:2006sv,Bellazzini:2020cot,Tolley:2020gtv,Caron-Huot:2020cmc,Arkani-Hamed:2020blm,Caron-Huot:2021rmr} (see also~\cite{Mizera:2023tfe} for a recent review). These results, non-perturbative in nature, help elucidate what constraints a low-energy effective theory has to satisfy in order to  admit a UV completion obeying certain cherished physical properties (locality, causality, unitarity). Perhaps more interestingly, this viewpoint can be turned around, so that observing a violation of such constraints at low energies would teach us that the fundamental laws of nature in fact {\it do not} obey one or more of these standard physical properties.

The extension of these considerations to theories with (spontaneously) broken Lorentz invariance is an interesting open problem, with potential applications ranging from early universe cosmology, to condensed matter physics, to high density nuclear matter. An interesting step in this direction has been taken in~\cite{Creminelli:2022onn}, where positivity bounds on current-current correlators were derived. As for scattering amplitudes the question is still unsettled. First of all,  defining scattering amplitudes in spacetime backgrounds that are not asymptotically flat or lack a time-like Killing vector, such as cosmological backgrounds, is problematic. Moreover, in the absence of boost invariance, physical excitations often correspond to metastable quasi-particles rather than exactly stable asymptotic states, i.e.~well defined single-particle states.\footnote{ A well studied example is that of phonons in superfluid helium-4, which have a decay rate $\Gamma \propto k^5$, where $k$ is their momentum. Being Goldstone bosons, their existence as single particle states is guaranteed for $k \to 0$ by the Goldstone theorem, but beyond that limit, there are no single particle states---just a continuum.}
Even setting aside the issue of properly defining an (approximate) scattering amplitude in these systems, it is far from clear what properties this object should satisfy. While unitarity and causality can be regarded as solid assumptions for a consistent $S$-matrix, crossing symmetry and analyticity cannot be taken for granted.\footnote{Analytic properties of momentum space correlators follow from causality and do not require Lorentz invariance. However, connecting correlators to amplitudes is a subtle step that can introduce new non-analyticities when Lorentz invariance is broken, as we will see below.}
In~\cite{Baumann:2015nta} the authors considered scattering processes in single-field inflationary models, and noticed the failure of crossing symmetry, but a complete analysis of the analytic structure of the amplitude at arbitrary energies is still missing.  Ref.~\cite{Grall:2021xxm} tried to extend the dispersive bounds on even $s$-derivatives of the forward scattering amplitude to theories with broken boost invariance. To do so, however, one needs to make assumptions on the UV behavior of the amplitude that appear to be not justified (see also~\cite{Aoki:2021ffc} for a critical discussion). There are also other approaches, for instance those of Refs.~\cite{Baumann:2019ghk,Pajer:2020wnj}.

Motivated by these difficulties, in this work we take a step back and analyze in some detail the tree-level $S$-matrix for
what we believe to be the simplest example of a UV complete relativistic theory with a translationally and rotationally invariant ground state that breaks Lorentz. The theory in question is that of a complex scalar field with a $U(1)$ symmetry and with renormalizable interactions only, at a nonzero chemical potential for the $U(1)$ charge. The ground state at nonzero chemical potential can be thought of as a relativistic superfluid, which defines a preferred rest frame and thus breaks Lorentz.
We shall analyze the analytic structure of the phonon scattering amplitude, and attempt to draw some general lessons. 

The paper is organized as follows.
In section \ref{sec:model} we describe the model, derive the dispersion relation and lay out
the method for computing scattering amplitudes.
As a first step, we compute the phonon decay rate in section \ref{phonondecay}.
We show that the decay rate (normalized by energy) has a maximum that scales not only
with the quartic coupling~$\lambda$, but also with the chemical potential $\mu$.
The phonon lifetime can thus be made long by lowering the chemical potential.
In section \ref{mainresults}, we give the $2 \to 2$ tree-level scattering amplitude in its
full generality, and study its various limits. Its positivity properties in the forward limit are
spelled out in \ref{sec:positivity}. In section \ref{beyond}, we attempt to draw
some general lessons, in particular about positivity in the limit of a small density or chemical potential.

\vspace{.2cm}

\noindent
\textbf{Note added:}
In the course of our investigation, we learned that the same model was being independently studied by
Creminelli, Delladio, Janssen, Longo, and Senatore.

\vspace{.2cm}

\noindent
\textbf{Conventions:}
When contracting spacetime indices, we adopt the mostly minus signature. Given a four-momentum of components $k^\nu$, we denote the four-vector by $k$, the energy $k^0$ by $\omega$, and the spatial momentum by the boldface $\k$. Even though Lorentz invariance is broken in our system, we use the standard relativistic normalization for single-particle states, $\langle \k | \k' \, \rangle = 2  \omega  \, (2 \pi)^3 \delta^{3}(\k-\k' \, )$.

We use latin $a, b, \dots =1,2$ indices from the beginning of the alphabet for components in field space, $\phi^a=(\phi^1, \phi^2)$. By $\epsilon_{ab}$ we denote the Levi-Civita tensor in field space. We adopt natural units: $\hbar=c=1$.

\section{A weakly coupled, UV-complete model for a relativistic superfluid}
\label{sec:model}

We consider the simplest UV-complete model for a relativistic superfluid, the $U(1)$-invariant theory of a complex scalar  field with quartic interactions at finite charge density. This model has been first studied in~\cite{Babichev:2018twg} and aspects of its one-loop dynamics at finite density have been extensively analysed in~\cite{Joyce:2022ydd,Nicolis:2023pye}. The benefit of this model as a theoretical laboratory is that it allows a weakly coupled description with analytic control across the whole range of energy and density scales, allowing to go beyond the low-energy effective field theory for the phonon dynamics.
In this work we shall be mainly concerned with the tree level dynamics of this model and we shall neglect higher-order or non-perturbative corrections.\footnote{In fact,  this model is well defined only perturbatively,  because of a Landau pole in the UV.
However, this model can always be UV completed at very high energies without modifying its dynamics in the energy regimes of interest. We shall not be concerned with these matters. The $(2+1)$D version of this model is well defined non-perturbatively and  shares the same tree-level dynamics as ours, up to phase space factors.}

The action is
\begin{equation}\label{eq:phi4model}
    S = \int d^4x \, \big[ \, |\partial \Phi|^2 + M^2|\Phi|^2 - \lambda |\Phi|^4 \, \big] \;  .
\end{equation}
Notice the sign of the mass term: we assume $M^2 > 0$, so that the internal $\U(1)$ symmetry is spontaneously broken, 
which leads to gapless Goldstone bosons already at zero density.
To study the finite charge density dynamics of this model, we introduce a chemical potential  $\mu$, which, without loss of generality, we can assume to be positive. As usual, this can be achieved by promoting the partial derivatives to ``covariant'' derivatives:
\begin{equation}
 \partial_\nu \Phi \longrightarrow   D_\nu \Phi  \equiv \big(\partial_\nu + i \, \xi_{\nu} \big) \Phi \; ,
\end{equation}
where $\xi_{\nu} = \mu \, \delta_\nu^{0}$.
Up to total derivatives, the finite $\mu$ action takes the form 
\begin{equation}
  \label{S2}
    S = \int d^4x \, \big[ \, |\partial \Phi |^2 +2i\, \Phi  (\xi \cdot \partial) \Phi^* + \big( \mu^2+M^2  \big) |\Phi|^2  - \lambda |\Phi|^4 \, \big] \; .
  \end{equation}
{Alternatively, one can derive the above from eq. (\ref{eq:phi4model}) by $\Phi \rightarrow e^{i\mu t} \Phi$.}
For what follows, it is convenient to write the complex scalar $\Phi$ in terms of real components as
\begin{equation}
    \Phi = \frac{1}{\sqrt{2}} \left( \varphi^1 + i \varphi^2  \right).
\end{equation}
Using a notation that is covariant in field space, our fields $\varphi^a =(\varphi^1, \varphi^2)$ at tree level have the  following expectation value on the finite-$\mu$ ground state $|\mu \rangle$:
\begin{equation} \label{vev}
v^a \equiv  \langle \mu | \, \varphi^a \, | \mu \rangle \; , \qquad   v = \sqrt{\frac{\mu^2+M^2}{\lambda }} \; ,
\end{equation}
where $v^2= v^a v^a$.
We can then expand the two real fields around their ground state as:
\begin{equation}
\varphi^a(x) = v^a + \phi^a(x).
\end{equation}
The action then takes the form
\begin{equation}
    S = S_{\rm free} + S_{\rm int},
\end{equation}
with
\begin{equation} \label{action}
\begin{split}
    &S_{\rm free} = \int d^4x \, \bigg[  \frac{1}{2}  \partial_\nu \phi^a \partial^\nu \phi^a  +  \epsilon_{ab}\,  \phi^a \left(\xi \cdot  \partial \right) \phi^b - \lambda\, v_a v_b \, \phi^a\phi^b  \bigg] \; ,\\
    &S_{\rm int} = - \int d^4x \bigg[ \dfrac{g_{abc}}{3!} \, \phi^a \phi^b \phi^c + \dfrac{\lambda_{abcd}}{4!} \,\phi^a \phi^b \phi^c \phi^d \bigg] \; ,
\end{split}
\end{equation}
where the couplings $g_{abc}$ and $\lambda_{abcd}$ can be taken to be totally symmetric,
\begin{equation} \label{couplings}
\begin{split}
    &g_{abc}=2\lambda \left(v_a \delta_{bc}+ v_b \delta_{ca}+ v_c \delta_{ab} \right),\\
    &\lambda_{abcd}= 2 \lambda \left(\delta_{ab} \delta_{cd}+\delta_{ac} \delta_{bd}+ \delta_{ad} \delta_{bc} \right).
\end{split}
\end{equation}
Working in Fourier space, the quadratic part of the action is 
\begin{equation}
S_{\rm free} = \dfrac{1}{2} \int \dfrac{\d^4 k}{(2\pi)^4} \,\phi^a\left(-k \right) K_{ab} \, \phi^b\left(k \right) ,
\end{equation}
where
\begin{equation}\label{eq:kinetic}
K_{ab}(k)= k^2 \delta_{ab} -2i \, \left(\xi \cdot k \right) \epsilon_{ab} - 2 \lambda \, v_a v_b.
\end{equation}
The Feynman propagators are easily obtained from the inverse of $K_{ab}$:
\begin{equation}\label{eq:propagator}
\begin{split}
&D^{ab}(k)\equiv i (K^{-1})^{ab}= \dfrac{i}{\det K} \Bigg(\Big(k^2 -2 (M^2+\mu^2)\Big)\delta^{ab} + 2i \, \left(\xi \cdot k \right) \epsilon^{ab}+ 2 \lambda \, v^a v^b \ \Bigg),\\
& \det K =  k^4 - 2 \left(M^2+\mu^2 \right) k^2 - 4 \left(\xi \cdot k \right)^2 \; , 
\end{split}
\end{equation}
where the $i \varepsilon$ terms are {kept implicit} (see \cite{Nicolis:2023pye} for a general treatment of these terms in this model).

We stress that,  for any non-zero value of the chemical potential $\mu$, the kinetic term has a momentum-dependent mixing term. As a consequence, it is not possible to diagonalize the quadratic action  by a {\it local} field redefinition. In order to keep locality manifest and deal with local fields only, we shall  work in the original, non-diagonal basis.

The dispersion relations of single particle excitations are easily obtained from the condition $\det K = 0$.
We obtain
\begin{equation}\label{eq:dispersion}
\omega^2_{\pm}(\k) = \k^2 + M^2 + 3 \mu^2 \pm \sqrt{4  \k^2 \mu^2 + (M^2 + 3 \mu^2)^2}.
\end{equation}
The minus sign corresponds to a gapless mode, since $\omega_{-} \to 0$ for $\k^2 \to 0$. This is nothing but the superfluid phonon, which for $\mu=0$ reduces to the ordinary, relativistic Goldstone boson of the spontaneously broken $\U(1)$ symmetry.  The zero-momentum group velocity of such a mode is the sound speed,
\begin{equation}\label{eq:cs}
c_s^2 = \Big(\frac{\d \omega_-}{\d \k}\Big)^2_{\k^2=0}= \dfrac{M^2 + \mu^2}{M^2 + 3\mu^2} \; .
\end{equation}
As we vary $\mu$ from zero to $\mu \gg M$, this interpolates between the speed of light, $c^2_s=1$, and the speed of sound in a conformal superfluid, $c_s^2 =1/3$.

The plus sign in \eqref{eq:dispersion} instead corresponds to a gapped mode, since $\omega_{+} \to \Delta > 0 $ for $\k^2 \to 0$, with
\begin{equation}\label{eq:gap}
\Delta^2 \equiv 2 M^2 + 6\mu^2.
\end{equation}
For $\mu = 0$, this is nothing but the squared  mass  of the radial mode of the relativistic $U(1)$ model in the spontaneously broken phase. For what follows, it will be sometime useful to use the gap $\Delta$ as the mass parameter of the model, rather than the original $M^2$, keeping in mind that the former depends on $\mu$ through \eqref{eq:gap}. And so, in particular, the low-density and high-density limits correspond to
\begin{align}
&\Delta^2 \to 2 M^2 \quad \mbox{for} \quad  \mu^2 \to 0 \; ,\\
&\Delta^2 \to 6 \mu^2 \quad \mbox{for} \quad  \mu^2 \to \infty \; .
\end{align}
Similarly, the sound speed \eqref{eq:cs} will appear as a parameter in our formulae, but it secretly is
\be
c_s^2 = 1- 4\frac{\mu^2}{\Delta^2}  \; ,
\ee 
with the standard $c_s^2 \to 1$ and $c_s^2 \to 1/3$ low-density and high-density limits. 
We can decide to keep $\Delta$ as the only dimensionful parameter, to which to compare energies and momenta, and trade $\mu$ for a combination of $c_s$ and $\Delta$,
\be
\mu^2 = \frac{1}{4} (1- c_s^2) \Delta^2 \; .
\ee
Similarly, the symmetry breaking scale \eqref{vev} is
\be
v^2 = \frac{c_s^2}{2\lambda} \Delta^2 \; .
\ee
Then, the low-density and high-density limit are simply expressed in terms of $c_s$ only
\begin{align}
\mbox{low densities ($\mu^2 \ll M^2$): }  & \quad  c^2_s \to 1 \; , \\
\mbox{high densities ($\mu^2 \gg M^2$): }  & \quad c^2_s \to 1/3 \; .
\end{align}
With this parametrization, the dispersion relations \eqref{eq:dispersion} take the form
\be
\omega^2_{\pm} (\k)= \k^2 + \frac12 \Delta^2\Big( 1 \pm  \sqrt{1 + 4(1-c_s^2){\k^2/\Delta^2}} \Big).
\ee

The two curves $\omega_{\pm} (\k^2)$ never cross, since their difference is manifestly positive. In this sense, there is a well defined ``light" excitation or particle and a well-defined ``heavy" one. The quotes are in order since, without Lorentz invariance, there is no invariant mass associated with our excitations. Also, 
we talk about two species of particles with arbitrary values for their momenta only as a matter of convenience, but it should be kept in mind that, since Lorentz boosts are broken, there is no symmetry relating particles of different momenta. From this viewpoint, different momenta correspond to different particle species \cite{Weinberg:1995mt} (see also \cite{Creminelli:2022onn} for further discussions about this and related points.)
With these caveats in mind, we shall refer to the lower branch as the ``gapless particle'', and to the upper one as the ``gapped particle''. 
Notice that the gap between the two particles asymptotes to a constant at high momenta: 
\begin{equation}
\omega_{+} - \omega_{-} \to 2  \mu \; ,\qquad {\rm for} \quad  \k^2\rightarrow \infty.
\end{equation}

Both dispersion relations are monotonically increasing functions of $\k^2$ and can be inverted to express $\k^2$ as a function of $\omega^2$:
\begin{equation}\label{eq:inverse_dispersion}
\k^2_{\pm}(\omega) = \omega^2 - \frac12  \Delta^2 \Big( c_s^2 \pm \sqrt{c_s^4 +4 (1-c_s^2) \omega^2/\Delta^2 } \Big) \; .
\end{equation}
We will need this formula in the following sections.

\subsection{$Z$ factors}
Another kinematic property of the model is that the field  normalization factors $Z$ (also known as wave-function normalization factors) are not constant, but depend on momentum. 
As usual, they are defined as the overlaps of the fields $\phi^a(x)$ with the single particle states $\ar$, where $\k$ is the momentum and $\alpha=\pm$ the particle species. We find it convenient to adopt the geometric viewpoint of~\cite{Cheung:2021yog}, and consider these factors as vielbein objects $Z^a_\alpha$, intertwining among the space of fields ($a$) and the space of states ($\alpha$). 
We define
\begin{equation}
 Z^a_{\alpha} (\k)= \langle \mu| \,  \phi^a(0) \, \ar \; , \qquad \qquad \bar Z^{a}_{\alpha}(\k) = \al \, \phi^a(0) \,  | \mu \rangle \; ,
\end{equation}
where $| \mu \rangle$ is the ground state at chemical potential $\mu$, latin indices run over field space, greek indices run over the space of states, and $\bar{Z}$ denotes complex conjugation.
Up to arbitrary momentum-independent phases\footnote{In the zero-density, Lorentz invariant limit, our choice of phases differs slightly from the more standard one, in which all $Z$ factors are real.}, 
for our model we find
\begin{equation} \label{e=A+B}
Z^{a}_{\alpha} (\k) = A_\alpha (\k) \, \hat v^a + i B_\alpha(\k) \, (\epsilon \cdot \hat v)^a    \; ,
\end{equation}
where $\hat{v}_a= v_a/v$, and $A$ and $B$ are the momentum- and species-dependent real factors
\begin{align}
A_\pm (\k) & = \frac{1}{\sqrt 2} \, \sqrt{1 \pm \dfrac{1}{\sqrt{1+4 (1-c_s^2) \k^2/\Delta^2}} } \; ,\\
B_\pm (\k) & = \pm\frac{1}{\sqrt 2} \, \sqrt{ 1 \pm \dfrac{(1-2 c_s^2)}{\sqrt{1+4 (1-c_s^2) \k^2/\Delta^2}}} \; .
\end{align}
Thanks to the model's $U(1)$ symmetry, all choices for  $\hat v_a$ are physically equivalent. We can keep it generic.

The $A$ and $B$ factors are functions of the particles' momentum $\k$ only.
It is of course possible to rewrite them as functions of the particles' energy $\omega$. Using the dispersion relations, after straightforward manipulations we obtain
\begin{align} 
A_\pm (\omega) & = \frac{1}{\sqrt 2} \,  \sqrt{1 \pm \dfrac{1}{ \sqrt{c_s^4  +4 (1-c_s^2) \omega^2/\Delta^2} \mp (1-c_s^2) } } \label{A of omega} \; ,\\
B_\pm (\omega) & = \pm \frac{1}{\sqrt 2} \,  \sqrt{ 1 \pm \dfrac{(1-2 c_s^2)}{ \sqrt{c_s^4  +4 (1-c_s^2) \omega^2/\Delta^2} \mp (1-c_s^2) }} \; . \label{B of omega}
\end{align}

In the following, we shall only be concerned with the gapless particles ($\alpha=-$). For these, it will be useful to know the leading low-energy and high-energy behaviors of our $Z$-factors:
\be \label{AB low energy}
A_-(\omega \to 0 ) \simeq \frac{\sqrt{1-c_s^2}}{c_s} \, \frac{\omega}{\Delta} \; , \qquad B_-(\omega \to 0 ) \simeq - {c_s}  \; ,
\ee
and
\be \label{AB high energy}
A_-(\omega \to \infty ) \simeq - B_-(\omega \to \infty ) \simeq \frac1{\sqrt 2} \; .
\ee
Notice that  for $\mu =0$ ($c_s^2 =1$) we have $A_ -= 0$ and   $B_- = 1$ at all energies, which is {\it not} the same as the $\mu \to 0$ limit of the high energy behaviors above. We thus see that the high energy and $\mu \to 0$ limits do not commute. This has important implications for the scattering amplitude, as we will see below.

An additional property that will be useful in what follows is that, as usual, near the poles the propagators take a simple form
\begin{equation}\label{eq:prop_pole}
D^{ab}(k) \approx \dfrac{i \, Z^{a}_{\alpha} (\k) \, \bar{Z}^{ b}_{\alpha}(\k)}{\omega^2 - \omega^2_{\alpha}(\k) + i \varepsilon}, \qquad\qquad  {\rm for} \quad \omega^2 \rightarrow \omega^2_{\alpha}(\k),
\end{equation}
as can be easily verified from the exact expression, but also, more readily, by inserting a complete set of states in the propagator and using standard polology manipulations (without Lorentz invariance).

The virtue of this simple model for a relativistic superfluid is that, by taking into account both the gapless and the gapped modes, we are able to describe scattering processes at arbitrary energies, without relying on a derivative expansion. This allows us to study the analytic structure of the (tree level) amplitude for arbitrary energies.

\subsection{Feynman rules and the LSZ reduction formula}

We denote time-ordered correlation functions in position space as 
\begin{equation}
G^{a^{}_1\dots a^{}_n}(x_1,\dots,x_n) = \zl T \phi^{a^{}_1}(x_1)\dots  \phi^{a^{}_n}(x_n) \zr .
\end{equation}
In momentum space we define
\begin{equation}
\tilde{G}^{a^{}_1\dots a^{}_n}(k_1,\dots,k_n) \,(2\pi)^4 \delta^{(4)}(k_1+\dots+k_n)= \int \prod_{j=1}^n \d^4x_j \,e^{+ i k_j  \cdot x_j}  G^{a^{}_1\dots a^{}_n}(x_1,\dots,x_n) .
\end{equation}
It is convenient to introduce the amputated correlation function, which we denote by $\mathring{G}$. Given a connected correlator $\tilde{G}_{(\rm conn)}$, the amputated correlator $\mathring{G}$ is defined through 
\begin{equation}\label{eq:amp_conn}
\tilde{G}_{(\rm conn)}^{ \, a^{}_1\dots a^{}_n}(k_1,\dots,k_n) = 
D^{a^{}_1a'_1}(k_1) \dots D^{a^{}_n a'_n}(k_n)\,\mathring{G}_{a'_1\dots a'_n}(k_1,\dots,k_n) \; ,
\end{equation}
where repeated indices are summed over.

For our action \eqref{action}, the Feynman rules for amputated vertices are simply
\vspace{0.5cm}
\begin{center}
\begin{tikzpicture}
\draw[black, thick] (-1.5,0) -- (0,0);
\draw (-1.5,0) node[anchor=east]{$a$};
\draw[black, thick] (0,0) -- (1,1);
\draw (1,1) node[anchor=west]{$b$};
\draw[black, thick] (0,0) -- (1,-1);
\draw (1,-1) node[anchor=west]{$c$};
\draw (2,0) node[anchor=west]{$= -i \,{g_{abc}} \;  ,$};
\end{tikzpicture}
\end{center}
\begin{center}
\begin{tikzpicture}
\draw[black, thick] (-1,1) -- (1,-1);
\draw (-1,1) node[anchor=east]{$a$};
\draw (1,-1) node[anchor=west]{$c$};
\draw[black, thick] (-1,-1) -- (1,1);
\draw (-1,-1) node[anchor=east]{$d$};
\draw (1,1) node[anchor=west]{$b$};
\draw (2,0) node[anchor=west]{\hspace{15pt}$= -i \, {\lambda_{abcd}} \; .$};
\end{tikzpicture}
\end{center}
Propagators, on the other hand, are given by eq.~\eqref{eq:propagator}:
\vspace{0.5cm}
\begin{center}
\begin{tikzpicture}
\draw[black, thick] (-2.6,0) -- (-0.2,0);
\draw (-2.6,0) node[anchor=east]{$a$};
\draw (-0.2,0) node[anchor=west]{$b$};
\draw (-2.6,0)-- node {\midarrow} (-.2,0);
\draw (1,0) node[anchor=west]{$= D^{ab}(k)$.};
\end{tikzpicture}
\end{center}

Assuming that there exist a set of asymptotic single-particle states, one can define the S-matrix. As we will see, without Lorentz invariance there are subtleties about this.
For the moment, we assume that the $S$-matrix exists. Taking as an example a $2 \to 2$ scattering process, the scattering amplitude $\M$ is then defined from the S-matrix through the relation
\begin{equation}
i\M_{\alpha\beta\rightarrow \gamma\delta} \, (2\pi)^4 \delta^{(4)}(k_1+k_2-k_3 - k_4) = \langle \gamma_{\k_3} \delta_{\k_4} \vert  S - 1 \vert \alpha_{\k_1} \beta_{\k_2} \rangle.
\end{equation}
The derivation of the LSZ reduction formula then goes through with minor modifications. Inserting  a complete set of asymptotic states in a 4-point correlation function, and going on-shell for the two incoming ($\alpha, \beta$) and the two outgoing ($\gamma, \delta$)  particles, we get
\begin{align}
& \tilde{G}^{abcd}(k_1,k_2,-k_3,-k_4)  \simeq \frac{i \bar{Z}^a_\alpha(\k_1)}{\omega_1^2 - \omega^2_{\alpha}(\k_1)} \dots  \frac{i {Z}^d_\delta(\k_4)}{\omega_4^2 - \omega^2_{\delta}(\k_4)} \times i\M_{\alpha\beta\rightarrow \gamma\delta} \\
& \mbox{for} \quad \omega_I^2  \rightarrow \omega^2_{\alpha_I}(\k_i), \nonumber 
\end{align}
where repeated particle-species indices are \emph{not} summed over, since they correspond to our particular choice of external states $\alpha, \dots, \delta$.
Using eqs.~\eqref{eq:amp_conn} and~\eqref{eq:prop_pole}, the LSZ reduction formula takes a particularly simple form once expressed in terms of amputated correlation functions:
\begin{align}
i\M_{\alpha\beta\rightarrow \gamma\delta} & = 
Z^{a'}_{\alpha} (\k_1)Z^{b'}_{\beta}(\k_2) \bar{Z}^{\, c'}_{\gamma} (\k_3)\bar{Z}^{\, d'}_{\delta} (\k_4)\,\mathring{G}_{a'b'c'd'}(k_1,k_2,-k_3,-k_4) \; , \label{eq:LSZreduced} 
\end{align}
where the repeated field-space indices are summed over.
We see immediately that the LSZ formula is in fact independent 
of the choice of interpolating fields.

\section{Phonon decay}
\label{phonondecay}

As a warmup computation, we analyze the decay of a gapless particle (phonon, for short) into two gapless ones. Given the form of our dispersion relations, this is kinematically allowed at all energies.\footnote{In particular, one can convince oneself that one needs a dispersion relation for the mode in question ($\alpha = -$, in our case) such that $\omega$ is a convex function of $|\k|$ but a concave function of $\k^2$. Our dispersion relation $\omega_- (\k)$ has these properties.}

The decay rate of a phonon into two is given by:
\begin{align}
\Gamma &= \frac{1}{2E} \int |\mathcal{M}|^2\text{d}\Pi_f \nonumber\\
&= \frac{1}{2E}  \times  \frac{1}{2} \int\frac{\text{d}^3\k_1}{(2\pi)^2 2\omega_{1}}\frac{\text{d}^3\k_2}{(2\pi)^2 2\omega_{2}} |\mathcal{M}|^2(2\pi)^4\delta^3(\p-\k_1-\k_2)\delta(E-\omega_{1}-\omega_{2}) \; ,
\end{align}
where we denote by $(E,\p)$ the four-momentum of the incoming particle, and by $(\omega_i,\k_i)$ the  four-momenta of the decay products. The decay amplitude $\mathcal{M}$ can be computed in complete analogy with eq. \eqref{eq:LSZreduced}. Given that the gapless mode is kinematically allowed to only decay into two gapless modes, we are interested in the following amplitude:
\begin{equation}
i\M = - i g_{abc} \, Z^{a}_{-} (\p) \bar{Z}^{\, b}_{-} (\k_1) \bar{Z}^{\, c}_{-}(\k_2) \; .
\end{equation}
Using the $Z$-factor parametrization \eqref{e=A+B} and the explicit form of the coupling constants \eqref{couplings}, we get
\be \label{decay amplitude}
i\M = -i \, 2\lambda v \Big[ A(\p)\big(3A(\k_1)A(\k_2)- B(\k_1)B(\k_2)\big)+
B(\p) \big( B(\k_1)A(\k_2) + A(\k_1)B(\k_2)\big) \Big] \;,
\ee
{where we have suppressed the ${}_-$ subscript label for phonons.}

As usual, one can simplify the phase space integral using the delta functions. Integrating over $\k_2$ and over the solid angle of $\k_1$, and changing the final integration variable from $|\k_1|$ to $\omega_1$ we find the following, still complicated, but somewhat more manageable form:
\begin{equation} \label{decay rate}
    \Gamma=\frac{1}{32 \pi E |\p|}\int_0^E \frac{\text{d}\k_1^2}{\text{d}\omega_1^2} \frac{\text{d}\k_2^2}{\text{d}\omega_2^2} \,  |\mathcal{M}|^2 \, \text{d}\omega_1,
\end{equation}
where it is understood that for both derivatives one should use the gapless $\k^2_- (\omega)$ dispersion relation, and that  $\omega_2$ is implicitly a function of $\omega_1$, $\omega_2 =  E - \omega_1$.

In \eqref{decay rate}, both the dispersion relations and the amplitude  \eqref{decay amplitude} are quite complicated functions of the integration variable. However, they smoothly interpolate between simple  low-energy and high-energy limits. The derivatives are of order of the inverse of squared propagation speeds\footnote{Notice that $d\omega^2/d \k^2 = v_{\rm ph} \, v_{\rm g} \, $, where $v_{\rm ph}$ is the phase velocity and $v_{\rm g}$ the group velocity.}, which  interpolate monotonically between $c_s^2 = 1/3$ and $1$. The $A$ and $B$ factors that make up the amplitude \eqref{decay amplitude}, interpolate monotonically between
\eqref{AB low energy} and \eqref{AB high energy}, that is between $A \sim \sqrt{1-c_s^2} \, \omega$, $B \sim 1$ at low energies and $A \simeq -B \sim 1 $ at high energies. Notice that $c_s ={\cal O} (1)$ in our model, and thus, for order-of-magnitude estimates, we do not need to keep track of it. On the other hand, $(1-c_s^2)$ can be much smaller than one---in the low-density limit---and so factors of it should be kept track of. 

In summary, even though we are not able to compute the full integral \eqref{decay rate} analytically, it is not difficult to estimate its order of magnitude. Let us neglect all ${\cal O}(1)$ numerical factors, including $c_s$. Barring cancellations, the amplitude \eqref{decay amplitude} is dominated by the $A BB$ term with the highest energy for the $A$ factor, which is $E$. That is
\be
{\cal M} \sim \lambda v A(E) \; , 
\ee 
and our rate \eqref{decay rate} is of order
\be
\Gamma \sim \frac{\lambda^2 v^2}{32\pi} \frac{A^2(E)}{|\p|} \; .
\ee
However, it so happens that at very low energies this contribution vanishes: there, each $A$ is proportional to its energy, while each $B$ is constant, and so energy conservation forces the three $A B B$ terms in \eqref{decay amplitude} to cancel. One is thus left with
\be
{\cal M} \sim \lambda v A^3(E) \; , \qquad E \to 0 \; ,
\ee
which leads to a much more suppressed rate:
\be
\Gamma \sim \frac{\lambda^2 v^2}{32\pi} \frac{A^6(E)}{|\p|} \propto E^5 \; , \qquad E \to 0 \; .
\ee

It is interesting to compare the decay rate to the incoming particle's energy, which is the oscillation frequency of the particle's wave-function. In particular,  the ratio
\be
\gamma(E) \equiv \frac{\Gamma(E)}{E} 
\ee
is the inverse of the quality factor: for $\gamma \ll 1$ the particle can undergo many oscillations before decaying; in that case it is a very narrow resonance. 
For our rates above, putting everything together, $\gamma$ interpolates between 
\be \label{low energy gamma estimate}
\gamma(E) \sim \frac{\lambda}{32\pi} (1-c_s^2)^3 \, \frac{E^4}{\Delta^4} \; , \qquad E \to 0 \; ,
\ee
and
\be \label{high energy gamma estimate}
\gamma(E) \sim \frac{\lambda}{32\pi}  \, \frac{\Delta^2}{E^2} \; , \qquad E \to \infty \; .
\ee
The transition between the two regimes, where $\gamma(E)$ develops a maximum, happens at $E_\star \sim \Delta/\sqrt{1-c_s^2}$, which corresponds to the {\it most unstable} our particle can be:
\be \label{gamma max estimate}
\gamma_{\rm max} \sim \gamma(E_\star) \sim \frac{\lambda}{32\pi} \, (1-c_s^2) \; .
\ee
The factor of $\lambda \ll 1$ is not surprising, since we are working in perturbation theory. What is perhaps less obvious is the extra suppression when $c_s^2$ is very close to one. Reinstating the chemical potential, we have
\be
E_\star\sim \frac{\Delta^2}{\mu} \; , \qquad \gamma_{\rm max}  \sim \frac{\lambda}{32\pi} \frac{\mu^2}{\Delta^2} \; .
\ee
We thus see that our particles become more and more stable when we lower the chemical potential: indeed, for vanishing $\mu$ we recover Lorentz invariance and our gapless particles become standard relativistic Goldstone bosons, which are absolutely stable.

\subsection{Limiting cases}
We can confirm our estimates above by  looking in detail at the low-energy and high-energy limiting cases. 

In the low-energy limit, $E \to 0$, all energies involved in the decay are small, and we have $\text{d}\k_-^2/\text{d}\omega^2\approx 1/c_s^2$ to lowest order in energies. Moreover, the $A$ and $B$ factors reduce to \eqref{AB low energy}, evaluated at the corresponding energies.
The decay rate \eqref{decay rate} then only requires a simple integration of a polynomial. The final expression, normalized to the decaying particle's energy, is
\begin{equation}
    \gamma =\frac{\Gamma}{E} \simeq \frac{3 \lambda  
   }{40 \pi } \frac{(1-c_s^2)^3 }{c_s^7}\frac{E^4}{\Delta^4} \; , \qquad E \to 0 \; .
\end{equation}
As a nontrivial check,   one can consider the same process in the low-energy phonon effective field theory. For our model  and to lowest order in derivatives, this takes a very simple form, ${\cal L} = P(X) = \frac{(X+M^2)^2}{4 \lambda}$ (see sect.~\ref{low energy section}). The phonon  decay rate computed in the effective theory matches exactly the result above (see also  \cite{Nicolis:2017eqo}).

For very high energies, the computation is also straightforward. In this limit the integrand in \eqref{decay rate} is dominated by the region where both $\omega_1$ and $\omega_2= E - \omega_1$ are large, so that no $A$ factor in \eqref{decay amplitude} is suppressed by its low-energy behavior \eqref{AB low energy}. This leaves out an integration region of fixed size $\delta \omega_1 \sim \Delta/ \sqrt{1-c_s^2}$, which is negligible compared to $E$ as $E\to \infty$. Then, inside the integral \eqref{decay rate} we can take $|\mathcal{M}|^2 \approx 8\lambda(M^2+\mu^2)$ and $\text{d}\k_-^2 / \text{d}\omega^2/\approx 1$. The  normalized decay rate becomes
\begin{equation}
     \gamma = \frac{\Gamma}{E}\simeq \frac{\lambda c^2_s}{8\pi}\frac{\Delta^2}{E^2} \; , \qquad E \to \infty \; ,
\end{equation}
which agrees with the estimate \eqref{high energy gamma estimate}.

\subsection{Full decay rate}
\label{fulldecayrate}
\begin{figure}[t]
\begin{center}
\includegraphics[width=9cm]{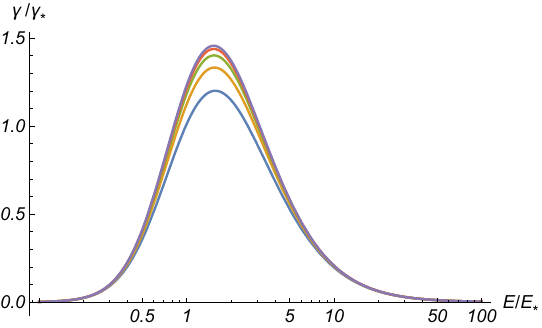}
\end{center}
\caption{\it \label{decay}The normalized decay rate $\gamma(E) = \Gamma(E)/E$ plotted against the incoming particle's energy $E$. The units $\gamma_\star$ and $E_\star$ are defined in \eqref{gammastar} and \eqref{Estar}. The different curves correspond to different values of $c_s^2$, from $c_s^2 = 1/3$ (bottom curve) to $c_s^2 \to 1$ (top curve).}
\end{figure}
In the intermediate regime where we are not able to evaluate the phase space integral analytically, we perform numerics to complement the asymptotic behaviors we just computed. Our results for these, suggest that we 
compute the normalized decay rate $\gamma (E) = \Gamma(E)/E$ in units of 
\be \label{gammastar}
\gamma_\star \equiv \frac{\lambda}{256\pi} \, \frac{(1-c_s^2)}{c_s} = \frac{\lambda}{64\pi} \frac{\mu^2}{c_s\Delta^2}
\ee
(the numerical prefactor is chosen for convenience),
as a function of $E$ measured in units~of 
\be \label{Estar}
E_\star \equiv  \frac{c_s^{3/2}}{\sqrt{1-c_s^2}} \Delta= \frac{c_s^{3/2}\Delta^2}{2 \mu} \; .
\ee
The result is displayed in fig.~\ref{decay} for different values of $c_s$, or, equivalently, for different values~of $\mu^2/\Delta^2$.

We checked that our numerical results match the correct asymptotics computed above. In particular, we confirm
that our particles become stable only asymptotically,  at zero or infinite energy. The normalized decay rate $\gamma$ peaks at energies $E \simeq 1.5 \times E_\star$ in any scenario, 
with a maximum of order $\gamma_\star$. And so, in conclusion, for any given $\lambda$ and $c_s$, the most unstable our particles can be is in the range 
\be
\underset{E }{\max} \, \frac{\Gamma}{E} \simeq (1.2 \text{---} 1.5) \times \frac{\lambda}{256\pi} \, \frac{(1-c_s^2)}{c_s} \; .
\ee

As emphasized above, the decay lifetime can be made arbitrary long by taking the small $\mu$ ($c_s \to 1$) limit. This partly motivates why we choose to study in detail the small $\mu$ limit, in some (though not all) of the discussions below.

\section{The scattering amplitude and its analytic structure}
\label{mainresults}

\subsection{Preliminary considerations}\label{preliminaries}
In a system that breaks Lorentz invariance, the freedom to change reference frame while keeping  the physics invariant is lost. As a consequence, the number of kinematic variables necessary to characterize a scattering process increases. For a $2\rightarrow 2$ scattering process, one starts with four independent momentum vectors, that is, twelve variables.
One can then use rotational invariance and energy-momentum conservation to remove seven combinations of these, since one can perform three independent rotations and impose four independent conservation constraints. The number of independent kinematical variables the amplitude can depend on is thus five.

A convenient choice, which allows us to connect to the Lorentz invariant case, is that of considering the usual Mandelstam invariants $s,t,u$, plus three similarly defined combinations of energies $\omega_s,\omega_t,\omega_u$:
\begin{equation}
\begin{split}
& s = (k_1+k_2)^2 ,\qquad \omega_s= (\omega_1+\omega_2), \\
& t = (k_1-k_3)^2 ,\qquad \omega_t= (\omega_1 - \omega_3), \\
& u = (k_1-k_4)^2 ,\qquad \omega_u= (\omega_1 - \omega_4) \; , 
\end{split}
\end{equation} 
where the $k_I$'s are the particles' momentum four-vectors, and the $\omega_I$'s the corresponding energies. 
From the definition of $s,t,u$ and energy-momentum conservation, it follows that these six quantities are not  independent, in agreement with our counting argument. For generic dispersion relations, there is a constraint of the form
\begin{equation}\label{eq:Mandelstam}
s+t+u = \sum_I \big(\omega_I^2 - \k_I^2) \; ,
\end{equation}
where each $\k_I^2$ has to be regarded as a function of $\omega_I$, and the $\omega_I$'s can be easily expressed in terms of $\omega_s,\omega_t,\omega_u$.

For scattering of identical particles, the functional form of the amplitudes as a function of the Mandelstam  and energy variables enjoys an extended form of crossing symmetry:
\be \label{extended crossing}
(s,\omega_s) \leftrightarrow (t,\omega_t) \leftrightarrow (u,\omega_u)  \; .
\ee 
This stems from the combinatorial relabeling symmetry that allows to reshuffle the external states. 
Two difficulties arise, though, in trying to use this symmetry to deduce dispersive bounds on low energy amplitudes, compared to the relativistic case. The first is that the number of invariants needed to parametrize a process is five, and complexifying all of them at the same time to derive dispersive bounds seems unfeasible. Choosing a kinematic configuration that reduces the number of independent invariants, for instance by specifying some values for $t$ or $\omega_s,\omega_t,\omega_u$, is possible, but in general breaks the extended crossing symmetry and leaves no residual symmetry relating $s$ and $u$.
The second is that the constraint~\eqref{eq:Mandelstam} is no longer a simple linear constraint, as it was for Lorentz invariant theories where $k^2=m^2$. 
And so, even if there existed a simple form of crossing symmetry relating $s$ and $u$, using \eqref{eq:Mandelstam} to express such a symmetry purely in terms of $s$ would make the symmetry quite complicated.
This is to be contrasted with the relativistic case, where crossing and the $s+t+u$ constraint imply a simple crossing symmetry  in the forward limit, $\M(s)^*=\M(4m^2 - s^*)$. 

This limitation might be overcome by choosing a more clever set of kinematic invariants. However, it seems difficult to do so in a model-independent way, i.e. independent of the precise functional form of the dispersion relation $\omega_{\pm}(\k)$. 
On the other hand, using standard Mandelstam invariants will allow us to easily compare our results to  familiar relativistic amplitudes. In particular, we expect to recover Lorentz-invariant results at $\mu=0$ and at very high energies, that is, when we can neglect the spontaneous breaking of Lorentz boosts. As we will see,  this comparison will be quite subtle.

To be concrete, we will only consider what is arguably the simplest process: the scattering of identical excitations in a center of mass configuration. By this we mean a configuration in which the incoming particles have opposite momenta of equal magnitudes, and as a consequence equal energies. In other words, the energies of all the incoming and outgoing single-particle states are equal and the spatial momenta are back-to-back: $\k_1=-\k_2$ and $\k_3=-\k_4$, with $\k_1^2= \dots = \k_4^2 \equiv \k^2$, $\omega_1 = \dots = \omega_4 \equiv \omega$. The amplitude can then be parametrized in terms of $s$, $t$, and $u$ only, since 
\begin{equation}\label{eq:omegaCM}
\omega_s^2 = 4\, \omega^2 =s, \qquad \omega_t=\omega_u=0 \;  . \qquad \qquad  \rm (center\; of \;mass)
\end{equation}

We notice that the center of mass value of $\omega_s$ is not a symmetric function of $s$, $t$, and $u$. As a consequence, the ordinary crossing symmetry $s \leftrightarrow t \leftrightarrow u$ of relativistic amplitudes  is \emph{broken} in all theories where the amplitude has an explicit dependence on $\omega_s$. 
Moreover, the relation~\eqref{eq:Mandelstam} takes now the form
\begin{equation}\label{eq:Mandelstam2}
s+t+u =  s - 4\, \k^2(s) \; ,
\end{equation}
which is also manifestly not symmetric under $s \leftrightarrow t \leftrightarrow u$.

Notice also a new form of non-analyticity in the complex $s$ plane: the relationship between $\k^2$ and $\omega^2$ in eq.~\eqref{eq:inverse_dispersion} has a cut. As a function of $s$, for gapless particles we have
\be \label{k(s)}
\k^2 (s)=\frac{1}{4}s - (M^2 + \mu^2) + \sqrt{\mu^2 s+ (M^2 + \mu^2)^2} \; ,
\ee
which has a cut at {\it unphysical} values of $s$, on the negative real axis, starting at $-\frac{(M^2 + \mu^2)^2}{\mu^2}$.
The same cut shows up in the $Z$-factors that we need in order to compute the amplitude---see eqs.~\eqref{A of omega} and \eqref{B of omega}.

\subsection{The full amplitude}
\label{sec:full}

The amplitude can now be computed straightforwardly using eq.~\eqref{eq:LSZreduced}.
At tree level there are only two independent Feynman diagram structures we need to consider: a four-point contact interaction, and two trilinear interactions connected by the exchange of an intermediate state  in the $s$, $t$, or $u$ channel. 
We obtain:
\begin{align}
i\M_{\alpha\beta\rightarrow \gamma\delta} & = - Z^{a}_{\alpha}(k_1) Z^{b}_{\beta}(k_2) \bar{Z}^{c}_{\gamma}(-k_3) \bar{Z}^{d}_{\delta}(-k_4) \times \\
& \Big[ i \lambda_{abcd} +
g_{abe} \, g_{cd e'} D_{e'e}(k_1+k_2) + g_{ace} \, g_{bd e'} D_{e'e}(k_1-k_3) + g_{ade} \, g_{bc e'} D_{e'e}(k_1-k_4) \Big] \; ,
\nonumber
\end{align}
where repeated indices are summed over.
The result is manifestly invariant under relabeling, as expected. 
In terms of the generalized Mandelstam invariants $s,t,u,\omega_s,\omega_t,\omega_u$, the amplitude satisfies the extended crossing symmetry \eqref{extended crossing}.

For the scattering of phonons (\emph{i.e.} gapless excitations, with $\alpha, \dots, \delta = -$) in a center of mass configuration,
we can  express the amplitude as a function of $s,t,u$ only by using eq.~\eqref{eq:omegaCM}. 
As previously remarked, this step is not symmetric under the exchange $s \leftrightarrow t \leftrightarrow u$, so that the resulting amplitude is not crossing symmetric in the usual sense. 
We do not assume the energy to be small, so that the gapped modes can appear as intermediate states.
The analytic structure of the amplitude is controlled by a new dimensionless combination
\begin{equation}\label{eq:y}
y \equiv \dfrac{\mu^2 s}{(M^2+\mu^2)^2} = \frac{1-c_s^2}{c_s^4} \frac{s}{\Delta^2}\; ,
\end{equation}
whose origin can be traced back to the $Z$ factors.

After some algebra, the phonon scattering amplitude can be expressed as 
\begin{equation}\label{eq:full_amplitude}
\M(s,t,u) =\lambda \left(1+ \dfrac{1-c_s^2}{c_s^2 \sqrt{1+y}} \right)^{-2} \Big( {\cal A}_1 + {\cal A}_2 + {\cal A}_3 +{\cal A}_4+ {\cal A}_5 \Big) ,
\end{equation}
where 
\begin{equation}\label{eq:full_amplitude_terms}
\begin{split}
{\cal A}_1 &= -\dfrac{1}{2} \dfrac{(1+\sqrt{1+y})^2}{1+y} \left[3+ \left(\dfrac{1}{s-\gap}  + \dfrac{1}{t-c_s^2\gap}+\dfrac{1}{u-c_s^2\gap} \right) c_s^2 \gap \right],\\
{\cal A}_2 &= -2 \dfrac{1+\sqrt{1+y}}{1+y} \, \dfrac{1-c_s^2}{s-\gap} \gap,\\
{\cal A}_3 &= \dfrac{y}{1+y} \left[\left(\dfrac{1}{s-\gap}  - \dfrac{1}{t-c_s^2\gap} - \dfrac{1}{u-c_s^2\gap} \right) 3 c_s^2 \gap - \dfrac{s-c_s^2\gap}{s(s-\gap)} \left(2c_s^2\gap\right) -1\right],\\
{\cal A}_4 &= 6 \dfrac{y}{(1+y)(1+\sqrt{1+y})} \,  \dfrac{1-c_s^2}{s-\gap} \gap,\\
{\cal A}_5 &= -\dfrac{3}{2} \dfrac{y^2}{(1+y)(1+\sqrt{1+y})^2}  \left[1+ \left(\dfrac{1}{s-\gap}  + \dfrac{1}{t-c_s^2\gap}+\dfrac{1}{u-c_s^2\gap} \right)3 c_s^2 \gap \right].\\
\end{split}
\end{equation}
The amplitude is manifestly symmetric under the exchange $t \leftrightarrow u$, but not under the exchange of either $t$ or $u$ with $s$. 
From the constraint \eqref{eq:Mandelstam2} and the inverse dispersion relation~\eqref{k(s)}, we have
\begin{equation}\label{eq:stu}
u = -t -s + 2 c_s^2 \gap(1  - \sqrt{1+y}),
\end{equation}
which we can use to obtain the amplitude as a function of $s$ and $t$ only, $\M(s,t)$.

\subsection{Analytic structure of the forward amplitude}

The amplitude of eq.~\eqref{eq:full_amplitude} has a rich analytic structure. We focus on the forward limit, $t=0$, which is well-defined and regular for the center of mass kinematic configurations we are considering.  We eliminate $u $ through \eqref{eq:stu}.

The first  feature  we notice is a branch cut, through the combination $\sqrt{1+y}$ ($y$ is defined in
eq. (\ref{eq:y})). This is remarkable, because such a branch cut is not enforced by unitarity and because it appears {\it at tree-level}, where we usually only have poles as non-analyticities. 
In the complex $s$ plane, the branch point is located at 
\begin{equation}\label{eq:s_star}
s_\star =- \dfrac{(M^2+\mu^2)^2}{\mu^2} \; .
\end{equation}
Close to it, the amplitude scales as ${\cal M} \sim (1+y)^{-1/2}$.
The combination $s_\star$ defines a new energy scale---in addition to the naive scales $\mu^2$ and $M^2$---controlling the analytic structure of the amplitude.
It is parametrically of the same order as $\mu^2$ for $M^2 \sim \mu^2$ or $\mu^2 \gg M^2$. 
On the other hand, it defines a genuinely new UV scale in the low density limit $\mu^2 \ll M^2$, and  plays an important  role as we shall discuss in the following sections.

The branch cut is directly related to the non-analyticity of the dispersion relations \eqref{eq:inverse_dispersion}.
Recalling that $s= 4 \omega^2$,  we see that the branch-point coincides with the (imaginary) value of energy where the two dispersion relations cross, in a way that is reminiscent of the analytic structure of the hydrodynamics expansion~\cite{Withers:2018srf,Grozdanov:2019kge,Grozdanov:2019uhi}.
Notice that, as usual, the location of the branch point is fixed, but there is freedom in choosing where to put the branch cut.
We choose to have it run along the negative real $s$ axis, from $s_\star$ to $-\infty$. Since its existence is not dictated by unitarity, the discontinuity across it does not seem to be associated with any unitarity cut. Singularities on the real positive $s$ axis, on the other hand, can be associated with the exchange of on-shell intermediate states, and their imaginary parts are dictated by unitarity through the optical theorem.

In addition to the square root branch cut, there are two poles on the first Riemann sheet, one in the $s$ channel and one in the $u$ channel:
\begin{equation}
\dfrac{1}{s-\gap} \qquad {\rm and} \qquad  \dfrac{1}{u-c_s^2\gap}.
\end{equation}
An additional pole is present on the second Riemann sheet {(where
$1 + \sqrt{1+y}$ vanishes)}, but it does not seem to have any immediate physical interpretation or relevance.
In the complex $s$ planes, the two poles above are at
\begin{equation} \label{poles}
s_1 =  \Delta^2 \qquad {\rm and} \qquad  
s_2 = -\Delta^2 \Big( 2 \sqrt{1-  c_s^2 + c_s^4} - (2-c_s^2) \Big),
\end{equation}
where we used eq.~\eqref{eq:stu} for the $u$-channel pole.
The $s$-channel pole occurs at $s_1>0$;  for the $u$-channel one, it is easy to see that for any value of $c_s$ 
\begin{equation}
s_\star < s_2 < 0 \; , \qquad \vert s_2 \vert < s_1 \; . 
\ee
Moreover, we have
\be
s_1 < |s_\star | \;  \qquad {\rm if \;and \;only \;if } \qquad  \dfrac{\mu^2}{M^2} = \dfrac{1-c_s^2}{3 c_s^2-1} < \frac{\sqrt{5}}{5}.
\ee

The amplitude is clearly not symmetric under inversion in the $s$-plane, $\M(s)\neq\M(-s)$. 
On the other hand, like standard  amplitudes in relativistic theories, it is polynomially bounded at infinity. In particular one can easily check that $\M(s)/s \rightarrow 0$ for $\vert s \vert \rightarrow \infty$. This agrees with the expectation that at very energies one should recover a standard relativistic amplitude, since the effects of Lorentz breaking should become less and less important at higher and higher energies.

We now consider some physically relevant limits.
\subsection{The low density limit}
\label{lowdensitylimit}
For $\mu=0$ ($c_s^2 = 1$) we recover the  relativistic amplitude for the $2\to 2$ scattering of Goldstone bosons in a Lorentz invariant abelian linear sigma model,
\begin{equation} \label{eq:relativistic_amplitude}
\M(s,t,u)\Bigg\vert_{\mu=0} = -6 \lambda -2  \lambda \left(\dfrac{2M^2}{s-2M^2} + \dfrac{2M^2}{t-2M^2} + \dfrac{2M^2}{u-2M^2}\right),
\end{equation}
which provides the first non-trivial check of our computation. In this limit only ${\cal A}_1$ contributes to the amplitude and, as expected, $c_s^2 \rightarrow 1$ and $s+t+u \to 0$. Moreover, the gap $\Delta$ becomes the mass of the radial mode, $\Delta^2 \to m_\rho^2 = 2M^2$.
For small but nonzero $\mu$, one can expand the amplitude in powers of $\mu$. Notice however that $\mu$ sometimes appears in the combination $y$ of eq.~\eqref{eq:y}. So, a small $\mu$ expansion at fixed $s$ corresponds not only to $\mu^2 \ll M^2$, but also to $\mu^2 s \ll M^4$, which at large $s$ is a stronger requirement.
In the forward limit ($t=0$), and after solving for $u$ with eq.~\eqref{eq:stu}, the leading order correction to \eqref{eq:relativistic_amplitude} is
\begin{equation}\label{eq:amp_mu2}
\M(s,t=0)\Bigg\vert_{\mathcal{O}(\mu^2)} = -\lambda\frac{4 \left(32 M^6 s^2+12 M^4 s^3-6 M^2 s^4-s^5\right)}{M^4 \left(2 M^2-s\right)^2 \left(2
   M^2+s\right)^2} \,\mu^2 \; .
\end{equation}
The apparent double poles at $s= \pm 2M^2$ are an artifact of the expansion of the physical single poles to order $\mu^2$, and are not physical. In other words, this expansion is only reliable away from the poles.

The low-energy expansion of this result in powers of $s$ presents interesting structural features. First,
due to the absence of inversion symmetry $s\to -s$, 
odd powers of $s$ appear, which is a novelty compared to the Lorentz invariant case. Second, all the coefficients of order $\mu^2$ in the  expansion in powers of $s$ are \emph{negative}:
\begin{equation}\label{eq:mu2neg}
\M(s,0)\Bigg\vert_{\mathcal{O}(\mu^2)} = \lambda \left(-\frac{8 s^2}{M^6} 
-\frac{3 s^3}{M^8}
-\frac{5 s^4}{2 M^{10}}
-\frac{5 s^5}{4 M^{12}}
-\frac{3 s^6}{4 M^{14}}
-\frac{7 s^7}{16 M^{16}}
-\frac{7 s^8}{32 M^{18}} 
-\dots \right)\mu^2.
\end{equation}
We shall postpone a detailed analysis of  positivity/negativity properties to section~\ref{sec:positivity}.

\subsection{The low energy limit}
\label{low energy section}
We can perform a non-trivial check of our results by considering the low-energy limit of the amplitude \eqref{eq:full_amplitude} at order $s^2$:
\begin{equation}\label{eq:s2}
\M(s,t=0)=\lambda\dfrac{2M^6+6\mu^2 M^4 + 9\mu^4 M^2 + 9 \mu^6}{2(M^2+\mu^2)^2 (M^2+3\mu^2)^3}\,s^2 + \mathcal{O}(s^3) \; .
\end{equation}
We can compare this  with the amplitude computed in the  phonon low-energy EFT, obtained by integrating out the gapped radial mode (see~\cite{Joyce:2022ydd} for a detailed derivation including one loop corrections). At tree level and at leading order in the derivative expansion we have
\begin{equation}
S_{\rm eff}[\theta] = \int \d^4x \,\dfrac{(X+M^2)^2}{4\lambda},  \qquad X= (\partial \theta )^2 \; ,
\end{equation}
where $\theta$ is the phase of our complex scalar $\Phi$. Expanding $\theta$ as 
\begin{equation}
\theta(x) = \mu \, t + \pi(x),
\end{equation}
where $\pi(x)$ denotes the phonon field, the action reduces to
\begin{equation}
S_{\rm eff}[\pi] =\int d^4x \,  \dfrac{1}{2 \lambda} \left[ \frac{\Delta^2}2 \big( \dot{\pi}^2 - c_s^2 (\nabla \pi)^2 \big)+  \frac{1}{2} \left( \partial_\nu \pi \partial^\nu \pi \right)^2 + {2\mu} \,\dot{\pi} \, \partial_\nu \pi \partial^\nu \pi + {\cal O}\big(\pi^5 \big)\right] \; . 
\end{equation}
Computing the $2\to 2$ scattering amplitude in the EFT for a center-of-mass kinematic configuration we find
\begin{equation}
\M_{\rm eff}(s,t,u) = \frac{8\lambda}{\Delta^4} \left[ \left(\frac{s}{2} -k^2  \right)^2 + \Big(\frac{t}{2} -k^2  \Big)^2 +\left(\frac{u}{2} -k^2  \right)^2 \right]  - 32 \lambda \frac{ \mu^2}{\Delta^6} \left(s - k^2 \right)^2 ,
\end{equation}
where
\begin{equation}
    k^2 = \left(1- \frac{1}{c_s^2}  \right) \frac{s}{4} \qquad {\rm and} \qquad     s+t+u = 4k^2.
\end{equation}
In the forward limit  and using eq.~\eqref{eq:cs} we recover the result of eq.~\eqref{eq:s2}.
This is the second non-trivial check of our computation, at low energy but arbitrary densities (or values~of $\mu$).

The coefficient of the $s^2$ term in the low energy amplitude \eqref{eq:s2} is manifestly positive, for arbitrary values of $\mu$. 
Due to the failure of crossing symmetry in the $s$ plane, terms with odd powers of $s$ appear in the low energy expansion, 
starting at cubic order: 
\begin{equation}
\M(s,0) \supset -\lambda\, \frac{\mu ^2 \left(12 M^8+58 M^6 \mu ^2+117 M^4 \mu ^4+126 M^2 \mu ^6+63 \mu ^8\right)}{4 \left(M^2+\mu
   ^2\right)^4 \left(M^2+3 \mu ^2\right)^4} s^3 \; . 
\end{equation}
Notice that the coefficient is manifestly {negative}, for arbitrary values of $\mu$.

Moreover, a new interesting structural feature emerges when considering the expansion of the low energy amplitude coefficients in powers of $\mu^2$ (with fixed $M^2$):
\begin{align}\label{eq:s2alt}
\M(s,0) = &\, \lambda \left(
\frac{1}{M^4} 
-\frac{8\mu^2}{M^6}
+\frac{93 \mu^4}{2 M^{8}}
-\frac{229 \mu^6}{M^{10}}
+\frac{1019 \mu^8}{M^{12}}
-\frac{4239 \mu^{10}}{M^{14}}
+ \dots \right)s^2 \\
& +  \lambda \left(
-\frac{3\mu^2}{M^8}
+\frac{67 \mu^4}{2 M^{10}}
-\frac{965 \mu^6}{4M^{12}}
+\frac{2821 \mu^8}{2M^{14}}
-\frac{29083 \mu^{10}}{4 M^{16}}
+ \dots \right) s^3 + {\cal O}(s^4) \nonumber
\end{align}
The coefficients of the series expansion in powers of $\mu^2$ have \emph{alternating signs}!

We shall generalize these statements to higher orders in $s$, and provide more details on the positivity and negativity properties of the low-energy expansion in section~\ref{sec:positivity}. 

\subsection{The high energy limit and a non-decoupling phenomenon}
\label{non-decoupling}

Given the appearance of a new energy scale $s_\star \sim \Delta^4/\mu^2$, the high energy limit of our amplitude is quite delicate. As we shall see, the limits $s\to \infty$ and $\mu^2\to 0$ (in $\Delta^2$~units) do not commute. The reason for this surprising behavior is the fact that the scale $s_\star$ runs away to (minus) infinity in the  $\mu^2\to 0$ limit.

Let us start by considering the forward amplitude ($t=0$). Taking the high energy limit $s\to \infty$, we find that the amplitude approaches the value $4\lambda$, for every fixed $\mu^2>0$. On the other hand, taking the high energy limit of the Lorentz invariant ($\mu=0$) result~\eqref{eq:relativistic_amplitude} for $t=0$, we find $-4\lambda$, which has the opposite sign!
To make the subtlety manifest, let us consider the case $\mu^2 \ll M^2$. In this case, the behavior of the amplitude can be characterized by three regimes: $s\ll M^2$ (low energy), $M^2 \ll s\ll M^4/\mu^2$ (high energy), and $M^4/\mu^2 \ll s$ (very high energy)---see fig.~\ref{fig:regimes}.

\begin{figure}[h]
\begin{center}
\includegraphics[width=5in]{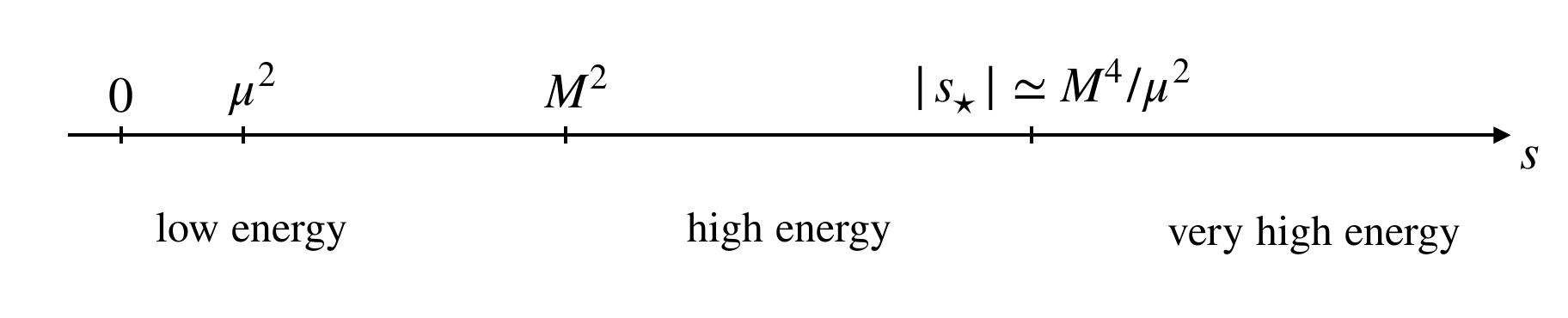}
\vspace{-15pt}
\end{center}
\caption{\label{fig:regimes} \it \small Energy scales and energy regimes in the small density limit, $\mu^2 \ll M^2$. The scale $s_\star$ correspond to the branch point on the negative $s$ axis, eq.~\eqref{eq:s_star}.}
\end{figure}

At high and very high energies, $s\gg M^2$, the forward amplitude $\M(s,0)$ simplifies significantly. Working at zeroth order in $M^2/s$ and $\mu^2/M^2$, but to all orders in $\mu^2 s /M^4$, we find 
\be
\M(s,0) \simeq 4\lambda - \dfrac{8\lambda}{\sqrt{1+\dfrac{\mu^2 s}{M^4}}} + \dots \qquad\qquad {\rm for}\quad s\gg M^2 \gg \mu^2.
\ee
The high energy forward amplitude interpolates smoothly between $-4\lambda$ in the ``high energy'' regime (corresponding to the $\mu=0$ expectation) and $4\lambda$ in the ``very high energy'' regime. The transition is governed by the branch point scale $\vert s_\star\vert$, as expected.
We checked that this simple expression provides an excellent approximation to the full analytic result~\eqref{eq:full_amplitude}, in its regime of validity. In going from negative to positive, the amplitude vanishes at a particular value of $s$ that we denote $s_0$. Including subleading corrections $\mu^2/M^2$, we find
\be
\M(s_0,0)=0, \qquad s_0 = 3\dfrac{M^4}{\mu^2} +2 M^2 + 10 \mu^2 + \dots,  \qquad\qquad {\rm for}\quad M^2 \gg \mu^2.
\ee

At face value, this behavior is quite surprising: why should the high energy amplitude be sensitive (\emph{at order one}!) to a tiny infrared $\mu^2$ deformation? In fact, the very high energy behavior of the amplitude is Lorentz invariant, as expected, but has an unexpected coefficient.
One might suspect  this peculiar behavior to be due to the large $s$ behavior of the forward amplitude's   being an intrinsically  UV/IR mixed observable, since $t\to0$. However, this tension persists and is in fact sharpened for genuinely hard scattering processes: those at fixed angle, corresponding to $t= b \cdot s$ with fixed nonzero $b$. We find that in the very high energy' regime $s\gg M^4/\mu^2 $, the amplitude approaches the constant value $-4\lambda$, instead of the expected $-6\lambda$ corresponding to $\mu=0$, which is instead attained in the intermediate high but not very high energy regime.

Also in this case, the limit $s\gg M^2$ of the amplitude simplifies significantly for small $\mu^2/M^2$. The hard scattering amplitude $\M(s,t=b\cdot s)$, at high energies and to all orders in $\mu^2 s /M^4$ is given by
\be
\M(s,b \cdot s) \simeq -6\lambda + \dfrac{2 s \mu^2}{M^4 +s \mu^2}\lambda+ \dots \qquad\qquad {\rm for}\quad s\gg M^2 \gg \mu^2, \;b\neq 0\;.
\ee
At leading order, the high energy amplitude for hard scattering is independent of $b$, for any fixed non-zero $b$. Also in this case the result interpolates smoothly between the expected $-6\lambda$, below the branch point scale $\vert s_\star\vert$, and $-4\lambda$ at infinite $s$.

This behaviour can be understood as follows: for $\mu=0$, our theory describes a system in which the internal $\U(1)$ symmetry is broken. The asymptotic one-particle states correspond to two scalar particles which are eigenstates of CP and have dispersion relations
\be 
\label{eq:high_disp}
\omega_{-} = \vert \k \vert, \qquad \omega_{+} \approx \vert \k \vert + \dfrac{M^2}{\vert \k \vert} \qquad\qquad {\rm for}\quad \vert \k \vert \gg M , \; \mu= 0\;.
\ee
On the other hand, the finite charge density phase $\mu \neq 0$ breaks CP as well as the internal $\U(1)$ symmetry, and the two effects compete in determining the asymptotic states. In the high energy regime $M^2 \ll s\ll M^4/\mu^2$, the $\U(1)$ breaking effects dominate over the CP breaking ones, and the asymptotic states are approximate eigenstates of CP with dispersion relations~\eqref{eq:high_disp} up to $\mu^2/M^2$ corrections. In this regime the finite density effects on the amplitude are small corrections to the $\mu=0$ result. In the very high energy regime, $s\gg M^4/\mu^2$, the CP breaking effects dominate and the asymptotic single particle states are approximate eigenstates of the $\U(1)$ charge with dispersion relations
\be 
\label{eq:veryhigh_disp}
\omega_{-} \approx \vert \k \vert + \mu, \qquad \omega_{+} \approx \vert \k \vert -\mu\qquad\qquad {\rm for}\quad \vert \k \vert \gg \dfrac{M^2}{\mu}, M \;.
\ee

Indeed, if one considers the scattering amplitude in the Lorentz invariant $\mu=0$ theory in the unbroken phase ($M^2<0$ in our notation) one can choose whether to take CP or $\U(1)$ eigenstates as asymptotic states, as both symmetries are unbroken. The first choice corresponds to the quanta of the $\phi^1$ and $\phi^2$ fields; the second to the quanta of $\Phi$ and $\Phi^*$. Since these states are all degenerate, one can choose whichever basis one wants. The two  choices however are not equivalent. For the first choice, the amplitude is $-6\lambda$, while for the second it is $-4\lambda$.
So, in going from low to high and then to very high energies, our gapless particles start off as phonons in a Lorentz-violating phase, become the relativistic quanta of the CP-odd field $\phi^2$, and eventually morph into the positively-charged quanta of the complex field $\Phi$.

The same conclusion can be reached by analyzing the kinetic matrix in momentum space~\eqref{eq:kinetic}, which in the small $\mu$ limit is
\be
K_{ab}(k)  \approx  k^2 \delta_{ab} -2i \, \left(\mu \cdot \omega \right) \epsilon_{ab} - 2 M^2 \, \hat{v}_a \hat{v}_b \qquad\qquad {\rm for}\quad \mu^2 \ll M^2 \;.
\ee
Since any vector is an eigenstate of $\delta_{ab}$,
the eigenstates of $K_{ab}$ are determined by the two competing terms $\left(\mu \cdot \omega \right) \epsilon_{ab}$ and $M^2 \, \hat{v}_a \hat{v}_b$. In the high energy regime the latter term, which is $U(1)$-violating but CP-preserving,  dominates. Conversely, in the very high energy regime, the former one dominates, and it is $U(1)$-preserving but CP-violating. All this is possible because when both symmetries are unbroken, we have exact degeneracy for our excitations.

The effect we just described constitutes an instance of \emph{non-decoupling} of $\mu$ at high energies.\footnote{Something qualitatively similar happens, for instance, in the familiar non-decoupling effects of Higgs physics. In those cases, however, the non-decoupling effects concern the low energy effective description. See e.g.~\cite{Falkowski:2019tft} for a recent treatment emphasizing the non-analytic origin of those effects.}
It provides also a physical interpretation for the scale $\vert s_\star \vert$ as the scale where a physical change of regime takes place: at $s> \vert s_\star \vert$, the CP breaking effects induced by the finite charge density dominate the dynamics over the effects induced by the $\U(1)$ symmetry breaking.

\subsection{Positivity properties of the forward amplitude}
\label{sec:positivity}

Consider our forward scattering amplitude, as a function of $s$ only: ${\cal M} = {\cal M}(s)$.

In a Lorentz invariant case, if we can expand a Goldstone tree-level scattering amplitude around $s=0$, we find
\begin{equation}
    \M (s)\Big\vert_{\mu=0} = a_2 s^2 + a_3 s^3 + a_4 s^4 + ... = \sum_{n=2}^{\infty} a_n s^n ,
\end{equation}
where the coefficients depend on the couplings and mass scales of the UV completion.
The crossing symmetry of Lorentz invariant amplitudes implies an $s \to -s$ inversion symmetry. As a consequence, for all integers $m \geq 0$,
\begin{equation}
    a_{2m+1} = 0 \; .
\end{equation}
From the tree-level positivity bounds, on the other hand, we have $\M^{(2m)}(0)>0 \,$. Therefore,
\begin{equation} \label{standard positivity}
    a_{2m} > 0 \; .
\end{equation}

Now, in our case Lorentz symmetry is broken by a nonzero chemical potential $\mu$. If we expand the Goldstone scattering amplitude $\M(s)$ around $s=0$, we find
\begin{equation}
  \label{Msexpansion}
    \M(s) = a_2(\mu) s^2 + a_3(\mu) s^3 + a_4(\mu) s^4 + ... = \sum_{n=2}^{\infty} a_n(\mu) s^n \;.
\end{equation}

In our explicit example, we find that for arbitrary values of $\mu$, the coefficients satisfy
\begin{equation}
  \label{Msexpansionsign}
    a_{2m}(\mu) > 0 \; , \qquad    a_{2m+1}(\mu) < 0 \; ,
\end{equation}
that is
\begin{equation}\label{eq:positivity1}
    (-1)^n   \M^{(n)}(0) > 0  \qquad {\rm for} \qquad n \geq 2 \; .
\end{equation}
In addition to the positivity bounds of the Lorentz invariant theory, we find that for our amplitude all odd derivatives (above $n=3$) are {\it nonzero} and {\it negative}.

\begin{figure}
\begin{center}
\hspace{50pt}\includegraphics[width=.6\textwidth]{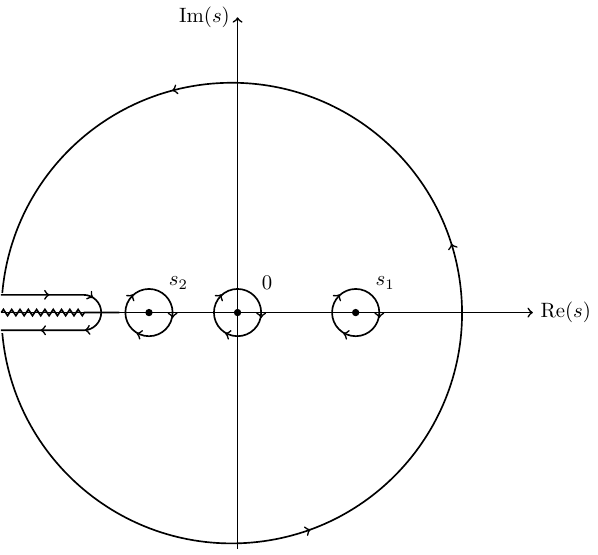}
\vspace{-15pt}
\end{center}
\caption{\label{fig:contour1} \it \small Analytic structure of $\M(s)/s^{n+1}$ for generic values of $\mu/M$, and contour of integration used to extract the low energy coefficient $a_n$, see eq.~\eqref{eq:an_Cauchy}. See section~\ref{sec:full} for a detailed discussion of the analytic structure of the forward amplitude.}
\end{figure}

The proof of this property for the amplitude~\eqref{eq:full_amplitude} is presented in Appendix~\ref{app:positivity}. This property is quite non-trivial from the point of view of the analytic structure of our amplitude. The amplitude is polynomially bounded, and so a contour argument in the complex $s$ plane can be used to evaluate the coefficients of the low energy expansion, see figure~\ref{fig:contour1}.
The presence of a branch cut that is not enforced by unitarity for negative values of $s$, however, complicates the analysis. 
We find
\be
\label{eq:an_Cauchy}
a_n  = - R_n(s_1) - R_n(s_2) + \chi_n,
\ee
where $R_n(s_{1,2})$ are the residues of the function $\M(s) /s^{n+1}$ at the poles \eqref{poles}, and $\chi_n$ is the branch cut contribution:
\be
\chi_n \equiv \dfrac{1}{2\pi i} \int_{-\infty}^{s_\star} \d s \, \dfrac{\mathcal{M}(s+ i\epsilon)-\mathcal{M}(s- i\epsilon)}{s^{n+1}}.
\ee
Notice that the discontinuity across the  cut is not related to any physical process, unlike in the case of unitarity cuts, since our cut is on the negative axis and $s \to -s$ is not a symmetry. In fact, the sign of $\chi_n$ is not fixed, and it goes from positive to negative as $\mu^2/M^2$ increases (we verified this numerically for $n=2,3$). Moreover, for large $\mu$ its size is of the same order as that of $R_n(s_1) + R_n(s_2)$, so that even if $R_n(s_1) + R_n(s_2)$ has fixed (negative) sign, the positivity statement~\eqref{eq:positivity1} crucially relies  on the value and sign of $\chi_n$, on which we do not have analytic control. The sign of $R_n(s_1)$ is fixed to be negative, by unitarity and the optical theorem~\cite{Adams:2006sv}. 
If $(-1)^{n+1} (R_n(s_1) + R_n(s_2)) > 0$ (which is true in our model for $n\geq 2$), then the positivity property~\eqref{eq:positivity1} corresponds to 
\be \label{chi_n condition}
\dfrac{\chi_n}{R_n(s_1) + R_n(s_2)} <1.
\ee
In figure~\ref{fig:branchcut_ratio} we plot the values of $\chi_n/(R_n(s_1) + R_n(s_2))$ for $n=2,3$ as a function of $\mu$. We see that $\chi_n$ does not have definite sign and its size relative to the sum of the residues can be of order one in absolute value. However, when this happens, $\chi_n$ and the sum of the residues have opposite signs, and thus \eqref{chi_n condition} is obeyed.
Notice that for small $\mu$ the branch cut contribution is suppressed due to the fact that $s_\star$ runs to $-\infty$ and the integrand becomes bounded by a smaller and smaller value. The positivity property satisfied by the amplitude appears to be highly non-trivial, as it relies on the sign and size of $\chi_n$. 

\begin{figure}
\begin{center}
\includegraphics[width=.46\textwidth]{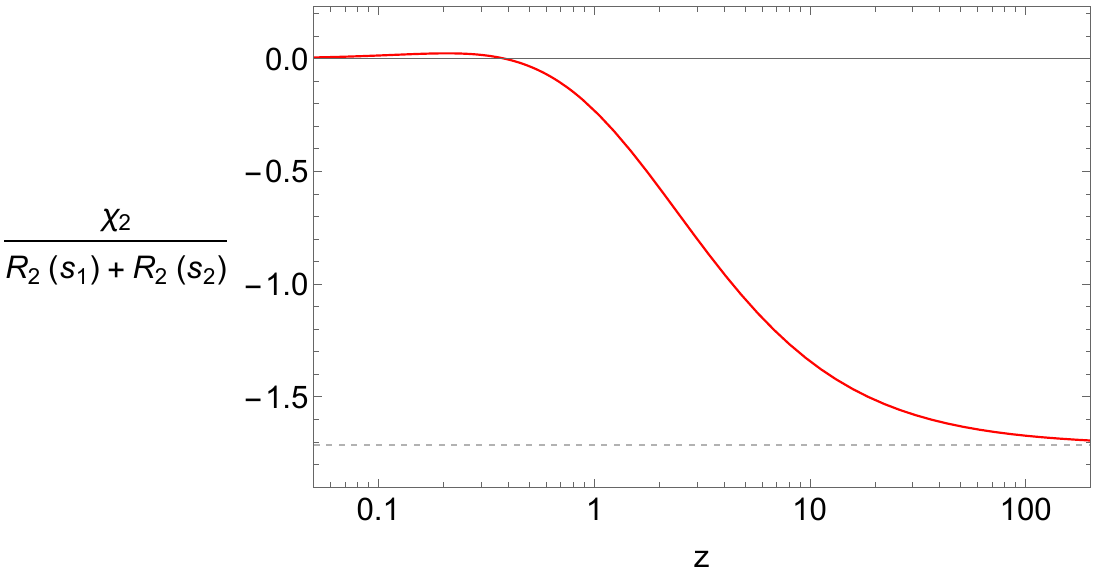}\quad
\includegraphics[width=.46\textwidth]{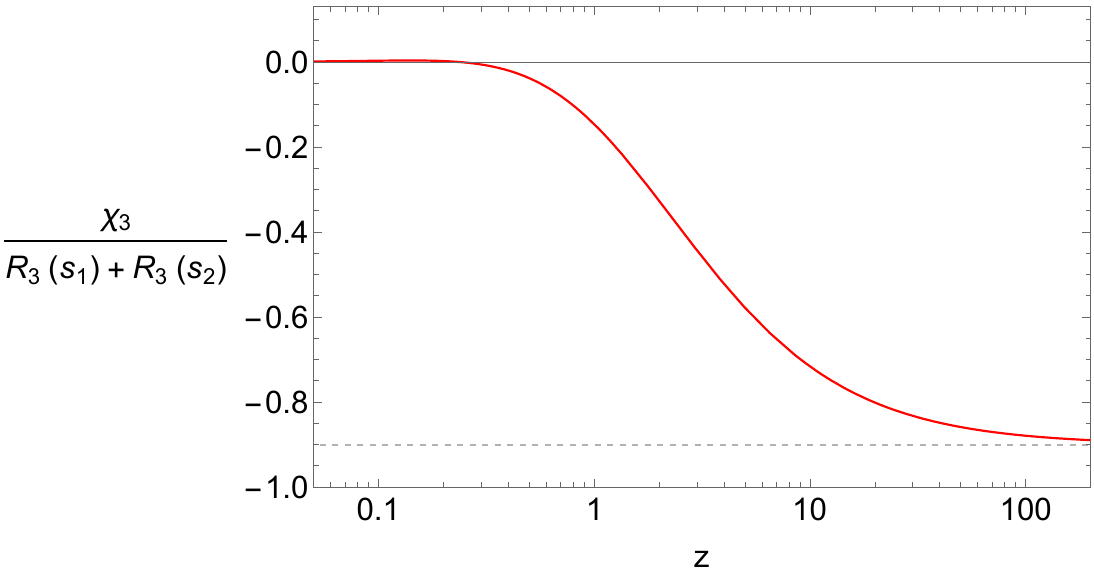}
\vspace{-15pt}
\end{center}
\caption{\label{fig:branchcut_ratio} \it \small The ratio defined in \eqref{chi_n condition} as a function of $z=\mu^2/M^2$, for $n=2$ (left) and $n=3$ (right). The quantity displayed depends only on $z$, and not on  $\lambda$ or $M$.}
\end{figure}

We do not know at the moment whether the positivity property~\eqref{eq:positivity1} has to be satisfied at arbitrary densities in a general relativistic QFT in a superfluid phase. 
In \ref{sec:negative_bound} we will try to understand what the needed assumptions are for it to be valid in the small density limit, at leading order in~$\mu^2$. 
Even if we do not have a proof for more general theories, we find  the generalized positivity property \eqref{eq:positivity1} of our amplitude quite suggestive.
In fact, there are additional remarkable properties that the amplitude~\eqref{eq:full_amplitude} satisfies in the forward limit, as we will now see.

Our coefficients $a_n(\mu)$, up to an overall factor of $1/M^{2n}$, are functions of $\mu^2/M^2$. Denoting $z\equiv \mu^2/M^2$, we find that the $a_n$ coefficients are meromorphic functions of $z$ in the complex $z$ plane, of the form 
\begin{equation}
a_n(z) =  (-1)^n \, \dfrac{P(z)}{Q(z)} \dfrac{\lambda}{M^{2n}}  ,
\end{equation}
where $P(z)$ is a polynomial of degree $2n-1$ with {\it positive integer} coefficients, and $Q(z)$ is a polynomial  of the form $q_n (z+1)^{2n-2} (3z+1)^{n+1}$, with {\it positive integer} $q_n$. In the  $\mu^2 \gg M^2$ (or $z\to \infty$) limit, we have $P(z)/Q(z) \sim 1/z^{n}$, as expected on dimensional grounds.

Additional properties emerge once we consider the double series expansion of the amplitude in powers of $\mu^2$ and $s$. 
We already noticed that the $s$-expansion of the ${\cal O}(\mu^2)$ amplitude has only negative coefficients, see eq~\eqref{eq:mu2neg}. On the other hand, we noticed that the $\mu^2$-expansion of the $s^2$ and $s^3$ coefficients of the amplitude has alternating signs, see eqs.~\eqref{eq:s2alt}. These properties are unified in the following, more general pattern. Denoting the double series expansion as
\begin{equation}
\M(s) = \sum_{n=2}^{\infty} \sum_{k=0}^{\infty} \, b_{k,n} \, z^{k} s^n \; ,
\end{equation}
where $z=\mu^2/M^2$, we have
\begin{equation}\label{eq:positivity2}
(-1)^{k}\,b_{k,n} >0 \; ,
\end{equation}
with the exception of $k=0$, for which all $b$ coefficients with odd $n$ vanish.
We can prove this property easily for small $n$ by contour integration in the complex $z$ plane. We spare the reader  the details.
However, as $n$ gets larger, the residue analysis becomes more and more intricate, and we have not found a simple way to prove~\eqref{eq:positivity2} for arbitrary $n$. We have checked this property with Mathematica for all coefficients up to order $\mu^{30} s^{30}$. 

Notice that the condition~\eqref{eq:positivity2} is independent of the positivity property stated in the previous section, which in terms of $b_{k,n}$ reads
\begin{equation}\label{eq:positivity3}
(-1)^{n}\, a_n >0 \; , \qquad a_n= \sum_{k=0}^{\infty} b_{k,n} \,z^{k} \; .
\end{equation}

\section{Beyond our example}
\label{beyond}
We now want to explore how much of what we have learned from our example can be extended
 to more general theories and physical situations. For simplicity,  we only consider theories in which Lorentz invariance is broken spontaneously rather than explicitly (notice that, as far as we know, this the case for all Lorentz-violating systems in nature). However, we will not use the fact that Lorentz invariance is non-linearly realized on the excitations of the system, and so in practice much of what we say can be applied to the explicit breaking case as well. We will also assume that there are gapless, derivatively coupled excitations, which we will call ``phonons". Upon minor modifications, one can give up this assumption as well. Finally, we will assume that Lorentz invariance is broken by the density of some conserved charge, or, equivalently, by a chemical potential. Again, this assumption will not be crucial, but it will provide us with a concrete framework. Presumably much of what we say can be extended to other forms of Lorentz violation. Finally, and this is instead an important assumption, we will only consider systems at zero temperature.

\subsection{General obstructions to defining an $S$-matrix}
The fact that our particles are unstable makes it impossible to sharply define an $S$-matrix: there are {\it no} single-particle asymptotic states that one can use to define the standard in- and out-states of $S$-matrix theory. This is a general problem for systems with broken Lorentz invariance: it stems from the fact that, without boost invariance acting linearly on excitations, there is no way to boost low-energy single-particle states into more energetic ones. And so, for instance, the existence of a stable particle at a given momentum, say $\k = 0$, does not guarantee the existence of stable particles with different momenta. 

This is a standard fact for phonons in superfluids and solids, even within the  low-energy EFT description: they have a decay rate $\Gamma \propto E^5$, and so they are absolutely stable only at zero energy. Moreover, the problem becomes even more general when we go to very high-energies, say much higher than the scale of the spontaneous breaking of Lorentz invariance: high-energy excitations will move close to the speed of light, which is higher than the propagation speed of whatever phonon-type excitations we have at low energies. Thus, high-energy excitations can Cherenkov emit low-energy ones. This process is a form of instability---it gives a width to the high-energy excitations.

This is to be contrasted with the standard case of a scattering process for particles moving in a Poincar\'e invariant vacuum. Consider for instance pion-pion scattering in QCD. There, although we are not able to {\it compute} the $S$-matrix at arbitrary energies, we can still {\it define} it, simply because a very energetic pion is nothing but the very boosted version of a low-energy one. Boost invariance allows us to define asymptotic states of arbitrary energy just by knowing the low-energy spectrum of the theory. Once an $S$-matrix is defined in this way, one can ask what analytic and positivity properties it has.

Going back to systems with broken Lorentz invariance, is everything lost? Clearly, if instabilities are slow enough, one can still talk about approximate asymptotic states with which to define an approximate $S$-matrix. For example, neutrons are unstable, but with an extremely slow decay rate compared to their mass, and so we routinely talk about neutron scattering (for example, for liquid helium itself!) The question is to what level of precision one can define and thus analyze the $S$-matrix. 

It seems to us that for particles that, like ours, admit $1 \to 2$ decay processes at tree level, the $S$-matrix can only be defined at tree level. The reason is that already at one loop there will be traces of the fact that our asymptotic states are not completely well defined. For example, one can put a loop on one of the external legs, schematically of the form $1 \to 2 \to 1$, which will then yield a nonzero width for that particle at this order in perturbation theory. 

For this reason, in our analysis of the scattering amplitude we only considered its tree level structure. To this order, the $S$-matrix is well defined, and we can study its analyticity and positivity properties, with the hope of understanding something general and useful. Notice however that this is only possible thanks to the fact that our UV-completion is weakly coupled. In other cases,  when the UV-completion is strongly coupled, for instance for superfluid helium-4, excitations are metastable  only at very low-energies, and truncating the $S$-matrix at tree level is not a good approximation to the physics of the system at higher energies.

Another possibility, which we have not explored, is to exploit the fact that, as emphasized in  sect.~\ref{phonondecay}, our particles formally become more and more stable at small densities, when $\mu^2 \to 0$. This is not surprising because the density (or $\mu^2$) plays the role of an order parameter for Lorentz breaking, and the obstructions we are emphasizing in this section are associated with the spontaneous breaking of Lorentz invariance. It could be that there is a way to push perturbation theory well beyond tree level, or even to approach the question non-perturbatively, but working instead to lowest non-trivial order in the density or the Lorentz breaking scale.

Finally, there is a more exotic possibility, about which we have nothing concrete to say at the moment. Phonons become more and more stable at lower and lower energies. So, there is a well-defined $S$-matrix for $E \to 0$. At very high energies there will be other particles, perhaps interpolated by the same local operators that interpolate the low-energy phonons. {\it If}, like in our case, these high-energy particles become more and more stable at higher and higher energies, then there is a well-defined $S$-matrix for $E\to \infty$ as well. It might be possible that there is a single function of the kinematical variables that on the one hand reduces to these scattering amplitudes in the low-energy and high-energy limits, but that on the other hand has certain analytic and positivity properties following from causality, locality, and unitarity, which one can use to run dispersive arguments like in the Lorentz invariant case. In a sense, such a function would be an analytic continuation of the $S$-matrix beyond the regions where the latter is defined. Such an analytic continuation will not have the interpretation of an $S$-matrix at intermediate energies, but studying its analytic properties might still turn out to be useful. 

Notice that the authors of ref.~\cite{Creminelli:2022onn} wisely steered away from $S$-matrix considerations: by considering the analytic structure of correlators of local operators that are well-defined at all scales, they bypassed completely the issue of defining asymptotic states. The only state they deal with is the ground state, which is exactly stable.

\subsection{Analytic properties of the forward scattering amplitude}

For forward scattering ($t=0$) of identical particles in a center of mass kinematic configuration, consider the tree-level amplitude as a function of $s$ only, ${\cal M}(s)$ (see sect.~\ref{preliminaries}). 
Given what we learned from our example, we expect the following properties:

\noindent \textbf{a) Existence of the forward amplitude.}
For gapless excitations with derivative interactions only and for center-of-mass kinematical configurations, the forward limit of the amplitude is finite.
On the other hand, away from center-of-mass kinematics, or with non-derivative interactions at low energies, new $1/t$ poles appear from $t$-channel diagrams, as noticed in~\cite{Baumann:2015nta,Nicolis:2017eqo}. However, this fact and the existence of a forward limit for center-of-mass kinematics can be checked directly in the low-energy effective theory, and so does not require assumptions about the behavior of the system at arbitrary energies. This is because $1/t$ poles are due to the exchange of gapless particles.

\noindent \textbf{b) No crossing symmetry.}
As  mentioned in sect.~\ref{preliminaries}, the general amplitude expressed as a function of all the Mandelstam and energy variables must enjoy an extended form of crossing symmetry, eq.~\eqref{extended crossing}.
However, once we express the amplitude as a function of $s$ and $u$ at fixed $t$, crossing symmetry in $s \leftrightarrow u$ is lost. Moreover,
the constraint \eqref{eq:Mandelstam2} at fixed $t$ gives quite a complicated relationship between $s$ and $u$ for generic dispersion relations. As a consequence, inversion in the $s$ plane is not a valid symmetry, $\M(s)\neq \M(-s)$.
However, we expect the amplitude to still satisfy a form of real analyticity, $\M^*(s)= \M(s^*)$.

\noindent \textbf{c) Unitarity.}
Unitarity, through the optical theorem, relates the imaginary part of the forward amplitude on the real positive axis to the total cross section.
As a consequence, singularities on the real positive axis have a physical meaning, which is preserved even with broken Lorentz invariance. At tree-level, singularities on the positive real axis consist exclusively of poles with $\Im \, \M(s)>0$, that is, with negative residues~\cite{Cutkosky:1960sp}.  

\noindent \textbf{d) Non-analyticities.}
The absence of inversion symmetry, $\M(s)\neq\M(-s)$, allows the appearance of new non-analyticities on the negative $s$ axis. Poles are expected, due to the $u$-channel exchange of gapped modes, but their location is now displaced by finite-$\mu$ effects compared to (minus) the location of the $s$-channel pole. In the explicit model that we studied, we had an $s$-channel pole  located at $s=\Delta^2$ (the gap of the radial mode~\eqref{eq:gap}), and a $u$-channel one  located at $u= c_s^2 \Delta^2$, suppressed by a sound speed factor. In terms of $s$, this pole was located on the negative real axis at a non-symmetric location.
Other singularities can appear on the negative $s$ axis, including branch cuts, {\it at tree level}, because of non-analyticities in the dispersion relations and in the field normalization factors at unphysical value of the energy.
In our case, such a non-analyticity was a branch cut, but we see no reason why in other cases we should not get new poles, for example.
No clear physical meaning appears to be associated with the discontinuity across our branch cut, in agreement with the fact that its presence is not dictated by unitarity.

\noindent \textbf{e) Polynomial boundedness.} At very high energies, or short distances, one expects to recover Lorentz invariance. This is similar to the familiar requirement for quantum field theory in curved spacetime that the short distance limit of correlators (in a well-behaved vacuum)  reduces to the flat space result. This is the case, for instance, for the Bunch-Davies vacuum in deSitter space. 
We thus expect $\M(s)$ to be polynomially bounded for $\vert s \vert \rightarrow \infty$, in particular $\M(s)/s^2 \rightarrow 0$ for $\vert s \vert \rightarrow \infty$. In fact, in our specific example, the tree-level amplitude goes to a constant at very high energies.
Non-analytic terms in the finite-$\mu$ amplitude, such as branch cuts, appear only at subleading order in the large $s$ limit, e.g. as $\mu/\sqrt{s}$.

\subsection{Towards dispersive bounds in the low density limit}
\label{sec:negative_bound}

We now consider a theory where Lorentz  invariance is only slightly violated. As mentioned above, we restrict to the case in which the breaking of Lorentz is measured by the chemical potential $\mu$ for a conserved charge, but we expect our considerations to be generalizable to other cases as well.
As before, we consider the tree-level forward amplitude ${\cal M}(s)$ for identical gapless modes in a center-of-mass kinematic configuration.

To make use of the properties of the Lorentz invariant  case, we assume some form of smoothness for the zero density limit:
We assume that the finite density amplitude is a smooth function of $\mu$ and that, in particular, poles and residues depend smoothly on $\mu$. We will use this assumption at finite values of $s$, and so the non-decoupling phenomenon of sect.~\ref{non-decoupling}, which happened for $s \to \infty$, will not be important. 

In theories for which the $\U(1)$ symmetry is spontaneously broken in the zero density phase, the zero density limit corresponds to  $\mu^2\to 0^+$. On the other hand, for theories with unbroken $\U(1)$ symmetry in the zero density phase, the zero density limit corresponds to $\mu^2\to (m^2)^+$, where $m$ is the mass of the lightest charged state~\cite{Nicolis:2023pye}. For the latter class our smoothness assumption is subtler, as there is a symmetry breaking phase-transition, which can induce non-analyticities.
We shall consider the former case only, and assume that the $\mu=0$ theory is in the spontaneously broken phase. We shall assume, moreover, that the $\mu=0$ amplitude satisfies the usual analyticity properties of relativistic amplitudes, and in particular $s \to -s$ crossing and polynomial boundedness.

\begin{figure}
\begin{center}
\hspace{50pt}\includegraphics[width=.6\textwidth]{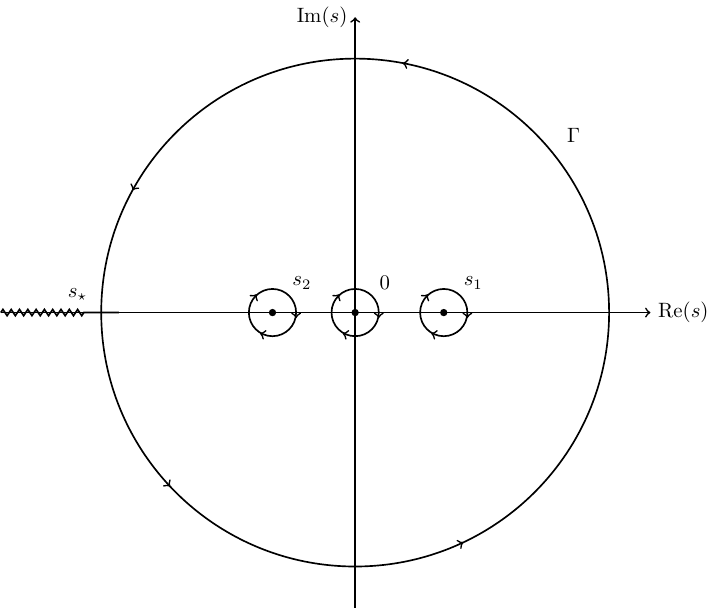}
\vspace{-15pt}
\end{center}
\caption{\label{fig:contour2} \it \small Integration contour in the limit of small $\mu/\Delta$. 
}
\end{figure}

For simplicity, let us assume that at vanishing $\mu$  the tree-level forward amplitude only has two poles, at $s = \pm \Delta^2$, that is, that there is only one massive state of  mass $\Delta$. Whatever we say can be straightforwardly generalized to cases with a richer spectrum. At small but nonzero $\mu$, we expect these two poles to shift by ${\cal O}(\mu^2)$, and we expect new analyticities---say, a branch cut, like in our example---to appear on the negative real axis starting at some $s_\star$ that goes to minus infinity as $\mu^2$ goes to zero, for instance $s_\star \sim -\frac{\Delta^4}{\mu^2} \ll -\Delta^2$. Our strategy is to consider a contour integral along a (counterclockwise) circle $\Gamma$ of radius $\vert s_\star\vert (1- \epsilon)$ that avoids the branch cut and is in the high energy region where the amplitude recovers a Lorentz invariant (polynomially bounded) value, see figure~\ref{fig:contour2}.\footnote{For the sake of the argument it is enough to consider $\epsilon \sim 1/2$. If the amplitude diverges at the branch point, as in our example, one should refrain from taking the  $\epsilon \to 0$ limit.} We are interested in capturing the leading $\mu^2$ corrections to the standard positivity bounds \eqref{standard positivity}, which enter at $\mathcal{O}(\mu^2)$. By Cauchy's residue theorem applied to the function $\M(s)/s^{n+1}$ we have 
\be\label{eq:residue_mu}
\dfrac{1}{2\pi i} \oint_\Gamma \d s  \,\dfrac{\M(s)}{s^{n+1}} = R_n(s_1) + R_n(s_2) + a_n,
\ee
where the residues correspond to the poles in $s_1$, $s_2$, and $0$ respectively.

The integral along $\Gamma$ can be bounded as follows:
\be 
\left\vert \dfrac{1}{2\pi i} \int_\Gamma \d s  \,\dfrac{\M(s)}{s^{n+1}} \right\vert \leq \max\limits_{\Gamma} \left\vert\dfrac{\M(s)}{ s^{n+1}}\right\vert \leq  \dfrac{C}{\vert s_\star \vert^{(n-1)}} \sim \frac{C}{\Delta^{2n-2}} \left(\dfrac{\mu^2}{\Delta^2}\right)^{n-1},
\ee
where  we used the smoothness hypothesis together with the polynomial boundedness assumption of relativistic amplitudes $\M(s)/s^2 \rightarrow 0$ for $\vert s \vert \rightarrow \infty$, and $C$ is a number of order of a small quartic coupling constant. We see that for $n\geq 3$ the contour integral is always parametrically suppressed compared to the $\mathcal{O}(\mu^2)$ terms we are interested in. We can thus consistently set it to zero.\footnote{The amplitude coefficient $a_2$ of order $n=2$ is generically non-vanishing already for $\mu=0$, and is constrained by the usual positivity bounds~\cite{Adams:2006sv}. We are mostly interested in the $a_n$ coefficients with odd $n \ge 3$. Therefore, the contribution from the $\Gamma$ contour is never expected to be important for small $\mu$.}

Let us now focus on the two residues $R_n(s_{1}),R_n(s_{2})$. They are
\be
R_n(s_{1,2}) =   \dfrac{{\rm Res} \big( \M(s_{1,2}) \big)}{s_{1,2}^{n+1}} \; .
\ee
To discuss their dependence on $\mu^2$, it is useful to introduce the residues of ${\cal M}$, defined by
\be
\M(s \to s_{1,2}) \simeq \dfrac{r_{1,2}(\mu^2)}{s-s_{1,2} (\mu^2)} \; .
\ee
In this equation we made explicit the $\mu^2$ dependence, which we are often keeping implicit. 
For $\mu = 0$, the amplitude has two poles related by crossing, at $s_{1,2} = \pm \Delta^2$, with residues
\be 
\bar{r} \equiv r_1(0) = - r_2(0) < 0 \; ,
\ee
where the sign follows from unitarity  \cite{Adams:2006sv}.
Working at order $\mu^2$, we obtain
\be
R_n(s_1) = \dfrac{\bar{r}}{\Delta^{2 n+2}} + \dfrac{1}{\Delta^{2 n+2}}\left[\dfrac{\d r_1}{\d \mu^2}  - (n+1) \dfrac{\bar{r}}{{\Delta^2}}  \dfrac{\d s_1}{\d \mu^2}\right]_{\mu=0} \mu^2 + \dots,
\ee
and similarly for $s_2$ (up to $n$-dependent signs due to crossing). As expected, the $\mathcal{O}(\mu^0)$ term survives in the sum $R_n(s_1) + R_n(s_2)$ only for even $n$, due to crossing. In this case, the positivity bounds of~\cite{Adams:2006sv} dictate positivity in the low density limit. Consider now the case of odd $n=2k-1$. 

 Using eq.~\eqref{eq:residue_mu} and putting everything together we  obtain 
\be\label{eq:res_mu_smallk}
a_{2k-1} = - \dfrac{1}{{\Delta}^{4k}} \left[ \dfrac{\d r_1}{\d \mu^2} + \dfrac{\d r_2}{\d \mu^2}  - 2k \dfrac{\bar{r}}{\Delta^2}  \left(\dfrac{\d s_1}{\d \mu^2}+ \dfrac{\d s_2}{\d \mu^2}\right)\right]_{\mu=0} \mu^2 + \dots
\ee
Notice that, apart from the overall prefactors, all terms are $k$-independent, with the exception of that multiplied by $2k$.
For large enough $k$, the sign of $a_{2k-1}$ is then controlled by the sign of the combination 
\be
 \bar{r} \left(\dfrac{\d s_1}{\d \mu^2}+ \dfrac{\d s_2}{\d \mu^2}\right)_{\mu=0} \; ,
\ee
where we dropped manifestly positive factors.

The sign of the residue $\bar{r}$ is fixed to be negative by unitarity.
We thus reach the conclusion that, under the physical and technical assumptions stated above, at small $\mu$ all even derivatives of the tree level amplitude are positive at $\mu = 0$,
\be
{\cal M}^{(2k)} (0) > 0 \; , \qquad \mu \ll \Delta \; ,
\ee
while all odd derivatives beyond a certain $k \sim \mbox{few}$ all have the same sign. In particular, 
\be
\label{oddsign}
{\rm sign} \, {\cal M}^{(2k-1)} (0)  = -{\rm sign} \left(\dfrac{\d s_1}{\d \mu^2}+ \dfrac{\d s_2}{\d \mu^2}\right)_{\mu=0} \; , \qquad \mu \ll \Delta \; .
\ee 

In our explicit example, the combination in parentheses happens to be positive. In fact, we checked that the same conclusion holds if we modify our model by a generic $\U(1)$ invariant potential, since the dispersion relation keeps a similar structure. However, we were not able to find a physical reason why that should be the case more in general. Notice that the pole at $s_1$ is physical, since it corresponds to a particle going on-shell in an $s$-channel exchange. And so, for example, one could envision that there exists a set of physical assumptions about the spectrum of the theory that  say something about the derivative $\d s_1/ \d \mu^2$. However, the pole at $s_2$ comes from a $u$-channel exchange, and thus corresponds to unphysical energies. In the Lorentz invariant case, crossing relates the two poles, but at the order in $\mu^2$ we are interested in, the $u \leftrightarrow s$ crossing symmetry is violated, and so it is not clear how to say anything about $\d s_2/\d \mu^2$ through physical assumptions. In our case, we had to use the explicit form of the inverse dispersion relation~\eqref{k(s)}, analytically continued to negative values of $s$, which is not obviously associated with any physical property of~theory.

A final remark is in order. The usefulness of positivity bounds like the ones derived above lies in their ability to constrain low-energy EFT's by postulating only a minimal set of physical assumptions about their UV completions. However, the relationship between the low-energy EFT and the low-energy behavior of the $2 \to 2$ amplitude can be complicated at high orders in $s$, especially in systems that break Lorentz invariance: in the EFT there are derivative corrections to the kinetic energies, to the trilinear interactions, and to the quadrilinear ones, and the number of independent structures escalates quickly at higher orders. So, for now we have only discussed bounds on the forward scattering amplitude. We leave understanding the constraints these put on low-energy EFT's for future work.

\section{Discussion}

In many ways, much of our knowledge about the analytic properties of
scattering amplitudes, and the associated positivity bounds, started from collecting data:
how do scattering amplitudes behave in a variety of theories?
This paper represents a modest step in collecting data for theories in which Lorentz invariance is
spontaneously broken. Specifically, we studied in detail the scattering of phonons in a superfluid, described by a
complex scalar with quartic interactions. At tree level, the $2 \to 2$ scattering
amplitude can be computed for any value of the chemical potential (which quantifies the amount of Lorentz breaking), at all energies and for arbitrary kinematic configurations.
The result is summarized by eqs. \eqref{eq:full_amplitude} and \eqref{eq:full_amplitude_terms}, which passes non-trivial checks in various limits (sections \ref{lowdensitylimit}, \ref{low energy section}, \ref{non-decoupling}). 

Focusing on the center-of-mass configuration and the forward limit, several features stand out:
1. Crossing symmetry in $s \leftrightarrow u$ is lost. As a result, when the amplitude is expressed as a function of $s$ alone, $\M(s) \neq {\cal M}(-s)$. In particular, the $s$ and $u$ poles are displaced from their Lorentz invariant locations by different amounts.
2. The tree-level
amplitude has a branch cut, which originates from
the non-analytic dispersion relation of the phonon. The branch point defines a new scale that goes
as the mass gap squared divided by the chemical potential $M^2/\mu$ (Figure \ref{fig:regimes}). 
3. The existence of the branch point implies the small chemical potential limit and the high energy limit do not commute. This can be traced back to a subtlety in the definition of the asymptotic states (section \ref{non-decoupling}). 
4. Derivatives of the scattering amplitude with respect to $s$ at $s=0$ follow an intriguing pattern: all even derivatives are positive and all odd derivatives are negative, for any value of the chemical potential (eqs. \eqref{Msexpansion} and \eqref{Msexpansionsign}). 

To understand better how this pattern emerges, we
investigate the limit of a small chemical potential $\mu$ (and fixed $s$, where there is no order of limit issue).
When $\mu=0$, Lorentz is unbroken, in which case it is well known all even $s$ derivatives are positive
and odd derivatives vanish (at $s=0$). 
When a small $\mu$ is turned on, we expect the even derivatives to receive small corrections but remain positive.
For the odd derivatives of a sufficiently high order, it can be shown that they must share the same sign, see eq.~\eqref{oddsign}. The sign is determined by how the $s$ and $u$ poles move in response to the turning on of the chemical potential.

A number of interesting issues are worth exploring.
For instance, we have chosen to stick with amplitudes at the tree-level, due to the instability of the phonon.
As shown in Section \ref{fulldecayrate},
the normalized phonon decay rate $\Gamma/ E$ has an upper bound that scales as $\lambda \mu^2$
where $\lambda$ is the quartic coupling. This scaling suggests, in the small chemical potential limit,
there's a meaningful way to interpret loop contributions that are $\lambda$ suppressed but not $\mu$ suppressed.
It would be useful to understand how to do so systematically.
Another interesting issue is the generality of the positivity/negativity statements for derivatives of the
scattering amplitude. For a small chemical potential, positivity of the even derivatives is simply a consequence of Lorentz invariance. Negativity of the odd derivatives, however, appears to rely on model dependent assumptions.
Can those assumptions be related to basic physical properties, such as the sound speed?
More ambitiously, is it possible to derive positivity/negativity statements for an arbitrary chemical potential?
Lastly, features such as the tree-level branch cut are a result of the non-analytic dispersion relation.
Is it possible to avoid them by considering appropriate correlation functions instead of scattering amplitudes?
We hope to address some of these issues in the near future.

\section*{Acknowledgements}

We thank Sergey Dubovsky, Sebastian Mizera, Riccardo Rattazzi and Petar Tadi\'c for useful discussions.
Our work is partly supported by the grant DOE DE-SC0011941.

\appendix

\section{Positivity properties of our forward amplitude}
\label{app:positivity}

We want to prove that, for $t=0$, our scattering amplitude ${\cal M}(s)$, see \eqref{eq:full_amplitude}, obeys infinitely many positivity properties on its derivatives:
\begin{equation}
    (-1)^n {\cal M}^{(n)} (s =0) > 0 \; .
\end{equation}
A function with these properties is called ``completely monotonic" (in the interval where the properties hold).\footnote{Strictly speaking, our function ${\cal M}$ and its first derivative vanish at $s=0$. We have checked that this subtlety does not spoil the arguments below, and so we will ignore it.} The version without alternating signs, with all pluses, defines instead an ``absolutely monotonic" function. Clearly, the two concepts are related by sending $s$~to~$-s$.
We will use repeatedly that sums and products of nonnegative completely monotonic functions are completely monotonic, and that compositions of completely or absolutely monotonic functions obey certain absolute/complete monotonicity properties~\cite{SRIVASTAVA20121649}.
These facts are obvious for absolutely monotonic functions, and a simple change of variables can extend them to completely monotonic ones.

It is convenient to use the variable
\begin{equation*}
    y = \frac{\mu^2 s}{(\mu^2+M^2)^2} \; ,
\end{equation*}
which is just a rescaling of $s$. So, we want to study the derivatives of the function ${\cal M}(y)$ at $y=0$.
Moreover, since in the expression of the scattering amplitude, the term $\sqrt{1+y}$ appears several times, we define
\begin{equation}
    r(y) = \sqrt{1+y} \; .
\end{equation}
The $n$-th derivative of $r(y)$ is 
\begin{equation}
    \frac{d^n r(y)}{dy^n} = (-1)^{n+1} \frac{(2n-3)!!}{2^n} (1+y)^{- \frac{2n-1}{2}} \; ,
\end{equation}
which means that $-r(y)$ is completely monotonic at $y=0$. Then, we can just study the derivatives of ${\cal M}(r)$ at $r=1$. If these obey the complete monotonicity condition, so do those with respect to $y$ at $y=0$.

${\cal M}(r)$ can be written as
\begin{equation} \label{pole pole g}
    {\cal M}(r) = \frac{1}{4} \frac{1}{\left(r +a -1   \right)^2} \frac{1}{(r +1)^2} \, g(r),
\end{equation}
where 
\begin{equation}
    a = \frac{3\mu^2 + M^2}{\mu^2 + M^2} = \frac{1}{c_s^2} \; , \qquad 1 < a \le 3,
\end{equation}
and the function $g$ is
\begin{equation} \label{g1}
    g(r) = \frac{b_1 r +c_1}{r^2 -a^2+a-1} + \frac{b_2 r + c_2}{r^2 + 2(a-1)r-a} + Q(r) \; , 
\end{equation}
where $Q(r)$ is a polynomial of $r$ of degree 4, and $b_1$, $b_2$, $c_1$, $c_2$ are $a$-dependent constants,
whose explicit expressions we spare the reader.
The first two factors in ${\cal M}(r)$ have the form 
\begin{equation}
    f(r) = \frac{1}{(r+b)^2} \; , \qquad b >0 \; ,
\end{equation}
and are thus positive and completely monotonic at $r=1$.

Then, what remains to be checked is that $g(r)$ is also nonnegative and completely monotonic at $r=1$.
We have checked explicitly $g(1)$ and its first four derivatives, and they all have the correct signs.
For higher derivatives, the polynomial $Q(r)$ does not contribute, and so we can focus on the first two terms in \eqref{g1}.
They have the general form
\begin{equation}
    h(r) = \frac{b \, r+c}{(r-r_+)(r-r_-)} \; ,
\end{equation}
where $r_\pm$ are the roots of the denominator.

By splitting $h(r)$ as the sum of two simple poles, and computing its derivatives at $r=1$, we find
\begin{equation} \label{hn}
    h^{(n)}(1) = \frac{(-1)^n n!}{r_+ - r_-} \left[ \frac{b \, r_+ + c}{(1-r_+)^{n+1} } 
    -\frac{ b \, r_- +c}{(1-r_-)^{n+1}  }  \right].
\end{equation}
For the two fractions in \eqref{g1}, we have
\be
b \to b_1 \; , \qquad c \to c_1\; , \qquad r_\pm \to r_{1\, \pm} \equiv \pm \sqrt{a^2-a+1} \;,
\ee
and
\be
b \to b_2 \; , \qquad c  \to c_2 \; , \qquad r_\pm \to r_{2 \, \pm} \equiv -(a-1) \pm \sqrt{a^2-a+1} \; ,
\ee
where the square root is real for all allowed values of $a$. Given the form of the derivatives~\eqref{hn}, for very large $n$
the $n$-th derivative of $g$ in $r =1$ will be dominated by the pole closest to $1$, which, for all allowed values of $a$, is $r_{2 , +}$:
\be
g^{(n \gg 1)}(1) \simeq \frac{(-1)^n n!}{(r_{2 , +} - r_{2, -})} \,  \frac{b_2 \, r_{2, +} + c_2}{(1-r_{2, +})^{n+1} } \; .
\ee
All factors on the r.h.s.~are positive, for all values of $a$, apart from the explicit $(-1)^n$. We thus see that, at high enough $n$, all derivatives of $g$ have the right sign for complete monotonicity.

How high is high enough? In lowering $n$, the first correction to the above approximate expression comes from the pole that is the second closest to one, which, again for all values of $a$, happens to be $r_{1, +}$. This correction becomes comparable to the contribution from $r_{2, +}$ at an $n \simeq \bar n$ such that
\be
 \bigg |\frac{b_1 \, r_{1, +} + c_1}{(1-r_{1, +})^{\bar n+1} }  \bigg| >  \bigg|\frac{b_2 \, r_{2, +} + c_2}{(1-r_{2, +})^{\bar n+1} } \bigg|  \; ,
\ee
where we used that the value of $(r_+ - r_ -)$ is the same for the two pairs of poles. We have
\be
\bar n < \frac{\log \Big | \frac{b_2 \, r_{2, +} + c_2}{b_1 \, r_{1, +} + c_1} \Big | }
{\log \Big| \frac{1-r_{2, +}}{1-r_{1, +}} \Big|}-1 \; .
\ee
Maximizing the r.h.s.~over $a$, we find that the largest possible integer $\bar n$ obeying this inequality is 2, which is quite smaller than the smallest $n$ ($n=5$) that we needed to complete our proof. That is, all derivatives of $g(r)$ have the right signs. This result can be proven more rigorously, but we omit a detailed proof.

So, in conclusion, all factors in \eqref{pole pole g} are nonnegative and completely monotonic, and so is our amplitude.

\bibliographystyle{JHEP}  
\bibliography{biblio.bib}

\end{document}